\documentclass[10pt,letterpaper]{article}

\usepackage[top=0.85in,left=2.75in,footskip=0.75in]{geometry}

\usepackage{amsmath,amssymb}

\usepackage{changepage}

\usepackage{textcomp,marvosym}

\usepackage{cite}

\usepackage{nameref}
\usepackage{hyperref}

\usepackage[right]{lineno}

\usepackage[nopatch=eqnum]{microtype}
\DisableLigatures[f]{encoding = *, family = * }

\usepackage[table]{xcolor}

\usepackage{array}

\newcolumntype{+}{!{\vrule width 2pt}}

\newlength\savedwidth

\raggedright
\usepackage[aboveskip=1pt,labelfont=bf,labelsep=period,justification=raggedright,singlelinecheck=off]{caption}

\makeatletter
\renewcommand{\@biblabel}[1]{\quad#1.}
\makeatother

\usepackage{lastpage,fancyhdr,graphicx}
\usepackage{epstopdf}
\fancyheadoffset[L]{2.25in}
\fancyfootoffset[L]{2.25in}
\usepackage{tikz}
\usepackage{graphicx}%
\usepackage{multirow}%
\usepackage{amsmath,amssymb,amsfonts}%
\usepackage{amsthm}%
\usepackage{mathrsfs}%
\usepackage[title]{appendix}%
\usepackage{xcolor}%
\usepackage{textcomp}%
\usepackage{manyfoot}%
\usepackage{booktabs}%
\usepackage{algorithm}%
\usepackage{algorithmicx}%
\usepackage{algpseudocode}%
\usepackage{listings}%
\usepackage{placeins}
\usepackage{wrapfig}
\usepackage{cleveref}
\usepackage[title]{appendix}
\usepackage{threeparttable}

\usepackage[normalem]{ulem}
\usepackage[font=small,labelfont=bf]{caption}
\theoremstyle{plain}%
\newtheorem{definition}{Definition}%

\theoremstyle{definition}%
\newtheorem{remark}{Remark}%

\begin{document}

\vspace*{0.2in}


\begin{flushleft}
{\Large
\textbf\newline{Persistence diagrams as morphological signatures of cells:\\
A method to measure and compare cells within a population} 
}
\newline
\\
Yossi Bokor Bleile\textsuperscript{1,2*},
Pooja Yadav\textsuperscript{3},
Patrice Koehl\textsuperscript{4},
Florian Rehfeldt\textsuperscript{3}
\\
\bigskip
\textbf{1} Institute of Science and Technology Austria, Klosterneuburg, Austria
\\
\textbf{2} School of Mathematics and Statistics, The University of Sydney, Sydney, NSW, Australia
\\
\textbf{3} Experimental Physics I, University of Bayreuth, Bayreuth, Germany
\\
\textbf{4} Department of Computer Science, University of California, Davis, California, America

\bigskip

* \href{mailto:yossi.bokorbleile@ista.ac.at}{yossi.bokorbleile@ista.ac.at (YBB)}


\end{flushleft}
\section*{Abstract}
   Quantifying cell morphology is central to understanding cellular regulation, fate, and heterogeneity, yet conventional image-based analyses often struggle with diverse or irregular shapes. We present a computational framework that uses topological data analysis to characterise and compare single-cell morphologies from fluorescence microscopy. Each cell is represented by its contour together with the position of its nucleus, from which we construct a filtration based on a radial distance function and derive a persistence diagram encoding the shape’s topological evolution. The similarity between two cells is quantified using the 2-Wasserstein distance between their diagrams, yielding a shape distance we call the PH distance.
We apply this method to two representative experimental systems—primary human mesenchymal stem cells (hMSCs) and HeLa cells—and show that PH distances enable the detection of outliers in those systems, the identification of sub-populations, and the quantification of shape heterogeneity. We benchmark PH against three established contour-based distances (aspect ratio, Fourier descriptors, and elastic shape analysis) and show that PH offers better separation between cell types and greater robustness when clustering heterogeneous populations. Together, these results demonstrate that persistent-homology-based signatures provide a principled and sensitive approach for analysing cell morphology in settings where traditional geometric or image-based descriptors are insufficient.

\section*{Author summary}

    Cell shape carries important information about how cells grow, respond to their environment, and differ from one another. However, cell populations often contain substantial natural variability, which makes it difficult for traditional image-analysis methods to reliably compare their shapes. In this study, we introduce a new way to measure and compare cell morphology using a mathematical approach known as topological data analysis. Our method describes each cell by the outline of its membrane together with the position of its nucleus, and converts this information into a compact \emph{shape signature} that can be compared across thousands of cells.

We apply this approach to two widely studied systems—human mesenchymal stem cells and HeLa cells—and show that it can detect unusual cell shapes, reveal sub-populations within heterogeneous samples, and quantify how variable a cell population truly is. We also compare our method with several commonly used shape-comparison techniques and find that it provides clearer separation between distinct cell types and more reliable clustering. Overall, this framework offers a robust and interpretable way to analyse cell morphology and may help researchers better understand the diversity and behaviour of cells in biological experiments.


\section{Introduction}

    Cells are the basic unit of life. Understanding how they grow, divide, die, and change shape is of central importance in virtually all fields of cell biology, including immunology, cancer biology, pathology, tissue, and organ morphogenesis during development, as well as in many other areas of the life sciences. Of significance to most studies aimed at these understandings is the concept of shape. The shape of a cell is defined by the geometric constraints of the space it occupies and is determined by the external boundaries and positions of the internal components. The shape is the result of the mechanical balance of forces exerted on the cell membrane by intra-cellular components and the extra-cellular environment. It is a geometric property controlled by a variety of biochemical pathways. Cell biologists study in parallel the morphology of cells (their geometry) with the regulation mechanisms that modify this morphology. These studies benefit from recent advances in microscopy and image processing techniques. Current microscopes provide 2D images that make it possible to study cellular shapes, or more precisely 2D projections of cellular shapes. The question remains as to how to measure and compare these shapes. This paper focusses on a new technique for performing these analyses. 
    
    Our interest in 2D shape comparisons is initially motivated by a seminal paper by Engler et al. that demonstrated that the mechanical properties of the extracellular matrix dictate the differentiation of human mesenchymal stem cells (hMSCs) towards various lineages\cite{Engler2006}. While the up- and down-regulation of genes and transcription factors towards terminal differentiation take up to several days or even weeks, experiments focused on the first 24 hours of hMSCs after seeding on a substrate showed a significant impact of matrix rigidity on the structural formation of acto-myosin stress fibres. Those studies introduced an order parameter $S$ that could be used as an early morphological descriptor of mechano-directed stem cell differentiation \cite{Zemel2010,Zemel2010b}. 
 
    The analysis described above was based on the filamentous structure of the cytoskeleton and its pattern formation; instead, we focus on the global cell morphology, in particular, the outline of the cellular cortex in two dimensions. Our goal is to use this morphology to identify possible differences in a cell population. For example, hMSCs are primary cells, collected from the bone marrow of human individuals in contrast to HeLa cells, an immortalised cell line. This leads to an intrinsic diversity for the cell population that is additionally influenced  by potential sub-populations of bone marrow fibroblasts  (roughly $5\%$)  \cite{hmscs-costa,hmscs-phinney}. Our aim is to see if geometry alone allows us to identify those sub-populations within a sample of cells.
    
        A 2D shape is defined as a domain $D$ in $\mathbb{R}^2$, delimited by its boundary, $\partial D$, often referred to as the contour of $D$. In all our applications, we will take the contour to be a piecewise smooth or polygonal Jordan curve, that is, a simple closed curve in $\mathbb{R}^2$. There are multiple geometric representations of such 2D shapes, leading to different methods for their characterisations. We briefly review two such representations.
    
\begin{figure}[t]
        \begin{center}
         \includegraphics[width=\textwidth]{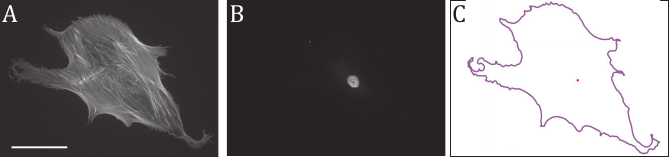}
           \caption{(A) Fluorescence microscopy image of a human mesenchymal stem cell (hMSC), scale bar represents 100 µm. (B) Fluorescence microscopy image of the corresponding nucleus. (C) Plot of the corresponding contour of that cell with the centre of the cell shown as a red dot.}
           \label{fig:fig1}
           \end{center}
    \end{figure}

    In the \textbf{{\em digital image}} representation, common to most real applications, raw data are provided in the form of 2D images (see Figure \ref{fig:fig1}A). In essence, the data to be understood and compared is a collection of pixels. Traditional methods of comparing such images usually proceed in three steps. They first define a set of well-chosen landmarks or key points on the surfaces of the shapes, then assign ``signatures'' to these key points (coordinates in a parameterising domain), and finally determine a map maximising the correspondence of signatures (for a review, see \cite{Chen:2021}). With the increase in computing power and the large number of image data sets that are generated, these ideas are often studied in the context of deep learning, where the key points and signatures are learnt from large data sets. Deep learning has become the predominant method used in 2D image analysis (see \cite{Zhou:2021} for a review of applications to medical image analysis). However, its applicability requires access to large data sets. In many cases, limited numbers of images are available, either because they are expensive to produce or because they model a rare phenomenon. This is the case for the stem cell images considered in this paper. In addition, deep learning remains something of a black-box procedure for classification. Cell biologists seek to understand the interplay between the geometry of a cell and the biochemical processes that are responsible for this geometry. They need a finer and more mechanistic understanding of the processes that drive shape, requiring mathematical approaches.
    
    A second representation of 2D shapes, which we refer to as \textbf{{\em shape as planar contour}}, is based on the curve describing the outer boundary of the shape (see Figure \ref{fig:fig1}C). This is well suited to applications focused on the geometric configuration of a shape, where factors such as the colour or grey level of the interior are either not relevant or not available. Methods for modelling the similarity between two shapes given as planar contours have been based on defining a distance between two curves in the plane. The proposed distances include the Hausdorff and Frechet distances \cite{Alt:2009}. Other techniques are based on the Poisson equation \cite{Gorelick:2006}, integral invariants \cite{Manay:2006}, and an elastic shape distance on the energy required to elastically deform one boundary contour to the other \cite{Joshi:2007, Srivastava:2011, Bauer:2014, Dogan:2015}.
    
It is worth providing a little more details on elastic energy as we will use it later for comparison with the new method introduced in this paper.
The elastic shape analysis framework, particularly through the Square-Root Velocity Function (SRVF) representation~\cite{Srivastava:2011}, provides a mathematically principled solution by treating shapes as points in an infinite-dimensional Riemannian manifold equipped with an elastic metric. This framework achieves invariance to translation, rotation, scale, and crucially, reparameterization, while allowing for elastic stretching and bending of curves~\cite{Kurtek:2012, Joshi:2007}. The SRVF representation transforms the computationally challenging elastic metric into the standard $\mathbb{L}^2$ metric, enabling efficient geodesic computation via path-straightening methods and facilitating statistical analyses including mean shape computation, principal component analysis, and shape classification~\cite{Cho:2019 ,Bharath:2020}. Recent extensions have broadened the applicability of this framework to medical imaging applications including tumor morphology analysis~\cite{Bharath:2018} and fiber tract comparison~\cite{Kurtek:2012}, demonstrating its versatility across diverse scientific domains.

    Paraphrasing a recent review paper by D. Chitwood and colleagues, ``Shape is data and data is shape'' \cite{Amezquita:2020}. As described above, shape is a signature of biological objects such as cells discussed above, that are significant for their biological functions. As such, shape characteristics are integral parts of the data that represent these biological objects. In contrast, there is a geometric structure within the data that is referred to as the shape of the data. Analysing the shape of data has become an essential section of data science, known as \emph{Topological Data Analysis}, or in short as TDA. TDA has its roots in the pioneering works of Robins \cite{Robins:2000}, Edelsbrunner et al. \cite{Edelsbrunner:2002} and Zomorodian and Carlsson \cite{Zomorodian:2004} in persistent homology and became popular with the publication of a landmark paper by Carlsson \cite{Carlsson:2009}. Since this paper was published, it has become ubiquitous in data science, with many applications in biology (see, for example, the review mentioned above \cite{Amezquita:2020} and the references therein illustrating applications in structural biology, evolution, cellular architecture and neurobiology). TDA is particularly useful when the data are represented in the form of a graph or network. As such, it proceeds by connecting data points to form a geometric complex structure whose topological behaviour is then used to analyse the data. Returning to the fact that the shape is data, a shape can be characterised through TDA using, for example, the Euler characteristic transform to study the morphology of barley seeds \cite{Amezquita:2022}, or the persistent homology transform to study the morphology of heel bones of primates \cite{pht}. 

    \FloatBarrier

    In this paper, we introduce a new method for analysing the morphology of a cell that falls into the second category described above, namely with the cell represented with its contour with an additional point $C$, taken to be the centre of mass of the cell nucleus. From TDA, we use \emph{persistent homology} to obtain a summary of the morphological features of the cell contour. We use the persistence of sub-level sets of the radial distance function from $C$ and compute the corresponding persistence diagram (see the next section for a primer on persistent homology applied to analysing cell contours). As the contour of each cell is a closed, non-self-intersecting curve, we know that it consists of a single connected component and a single $1$-cycle. These two cycles correspond to a persistent cycle with infinite life (called \emph{essential} cycles) in dimensions $0$ and $1$, respectively. We combine the information from these two persistent cycles by pairing the birth of the essential connected component with the birth of the essential $1$-cycle. A pair of cells is then compared by computing the \emph{2-Wasserstein distance}  between their \emph{persistence diagrams}, providing a measure of similarity between the two cells. We can then apply various clustering techniques to these similarity scores, to identify homogeneous populations of cells.
    
   The paper is organised as follows. The next section (\Cref{sec:theory}) introduces the concept of persistence homology  applied to analysing the morphology of a cell, the construction of the persistence diagram of a cell contour, and the computation of the Wasserstein distance between two persistence diagrams. The Materials and Methods section (\cref{sec:materials-and-methods}) gives information on the experimental data and implementations of the methods mentioned above. It includes a description of three standard methods for computing the distance between two curves (aspect ratio distance, Fourier descriptor distance, and elastic shape distance), as those methods will be compared to our persistence homology distance. The Results section (\cref{sec:results}) discusses the applications of our new method, PH, on two systems, hMSCs  that are expected to include sub-populations and, in contrast, HeLa cells whose morphologies are expected to vary continuously. An extensive analysis that compares PH with existing cell contour distances is included. We conclude with a discussion of how to integrate persistent homology with other more traditional techniques for comparing cell shapes.

\section{Theory: persistence homology applied to analysing cell contours}\label{sec:theory}

\subsection{Persistent Homology on Contours}\label{subsec:ph}

    Given a microscopy image of a fixed and immuno-stained cell, we use a graph $G$ to represent the boundary in $2$ dimensions. This graph is a list of ordered vertices (pixel locations), $V$, with edges, $E$, between neighbouring vertices. Note that $G$ is connected and that every vertex has degree $2$, so $G$ consists of precisely one $1$-cycle. We extract morphological information using the persistence of connected components of the sub-level sets of a radial function from the centroid of the nucleus. 
    
    For a graph $G$, we say that two vertices $v_1, v_2$ are in the same \emph{equivalence class}, or \emph{connected component}, if there is a path $\gamma$ from $v_1$ to $v_2$. For each connected component of $G$, we choose a representative vertex $v$ and denote the set of vertices $v'$ connected to $v$ by $[v]$. We call the set $\{ \, [v] \, \text{ for } v \in G\}$ the \emph{connected components} of $G$.
    
    To use persistent homology, we need to define a filtration on $G$.

    \begin{definition}[Sub-level sets and sequence of graphs]\label{def:sublevel}
        Let $f$ be a function from the vertices $V$ of a graph $G$ to $\mathbb{R}$, and fix $a \in \mathbb{R}$. The \emph{sublevel set} $G_a:= f^{-1}((-\infty, a])$ is the sub-graph consisting of the set $V_{a}$ of vertices $v$ with $f(v) \leq a$ and the set of edges $E_{a}$ between any pair of neighbouring vertices that are both in $V_{a}$. Note that for any \[a \leq b \in \mathbb{R}\] we have \[ f^{-1}((-\infty, a]) \subseteq f^{-1}((-\infty, b]), \] and the sub-level sets form a sequence of nested graphs. 
    \end{definition}
        
\subsection{Persistence Diagrams}\label{subsec:diagrams}

    Given a nested sequence of graphs $G_0 \subseteq G_1 \subseteq \ldots \subseteq G_{\alpha}$ (in general $G_\alpha = G$ the full graph), we can track the changes in the connected components of the graphs as the filtering parameter varies. Consider some $G_{\beta}$, and let $C_{\beta}:=\left \{\left [v_j\right ]^{\beta} \right \}_{j=1}^{n_i}$ be the set of connected components in $G_{\beta}$. For each connected component of $G_{\beta}$ we choose a canonical representative vertex, namely the vertex with the lowest function value. We say that a connected component $\left [ v_j \right ]$ is \emph{born} at time $\beta$ if there is no vertex $v_k$ in the connected component $\left[v_j\right]$ such that $v_k$ is in $C_{\beta - 1}$. We say $\left [ v_j \right ]$ \emph{dies} in $\gamma$ if in $G_{\gamma}$, $[v_j]$ becomes a path connected to a component born before $v_j$. For any pair $\beta \leq \gamma$ we obtain a map $\mathfrak{A}_{\beta}^{\gamma}: C_{\beta} \rightarrow C_{\gamma}$, which is induced by the inclusion $\iota_{\beta}^{\gamma}: G_{\beta} \rightarrow G_{\gamma}$. 

    \begin{remark}
        The map $\mathfrak{A}_{\beta}^{\gamma}: C_{\beta} \rightarrow C_{\gamma}$is obtained from the inclusion $\iota_{\beta}^{\gamma}: G_{\beta} \rightarrow G_{\gamma}$ by
        \begin{align*}
            \mathfrak{A}_{\beta}^{\gamma}\left( \left[ v \right] \right) := \left[ \iota_{\beta}^{\gamma}(v) \right],
        \end{align*}
        which is a well-defined map.
    \end{remark}

    The births and deaths of the connected components can be visualised in a \emph{persistence diagram}. 
    \begin{definition}[dimension $0$ Persistence Diagram]\label{def:persistencediagram}
        Let $f$ be a function from a graph $G$ to $\mathbb{R}$, and let $\mathfrak{G} = \{G_a\}_{a \in \mathbb{R}}$. Let $C = \bigcup_{a \in \mathbb{R}} C_a$ be the set of connected components across the sequence of graphs $\mathfrak{G}$. The \emph{dimension $)$persistence diagram}, $\mathfrak{D}(\mathfrak{G})$ of $\mathfrak{G}$ is the multi-set of points $(b_j,d_j) \in \mathbb{R}^2$, where $b_j$ is the birth time of $[v_j] \in C$, and $d_j$ its death time. A point with $d_j = \infty$ is called an \emph{essential point}, and the corresponding equivalence class an \emph{essential class}.
    \end{definition}

    \begin{remark}
        Here, we have only defined the persistence diagram in dimension $0$. It can defined analogously for in dimension $n$ be replacing connected components with $n$-cycles.
    \end{remark}
    
    We can also define these filtrations and persistence diagrams algebraically, including persistence modules, as in \cite{chazalpersmods}. 

        \begin{figure}[H]
        \begin{center}
            \includegraphics[height=12cm]{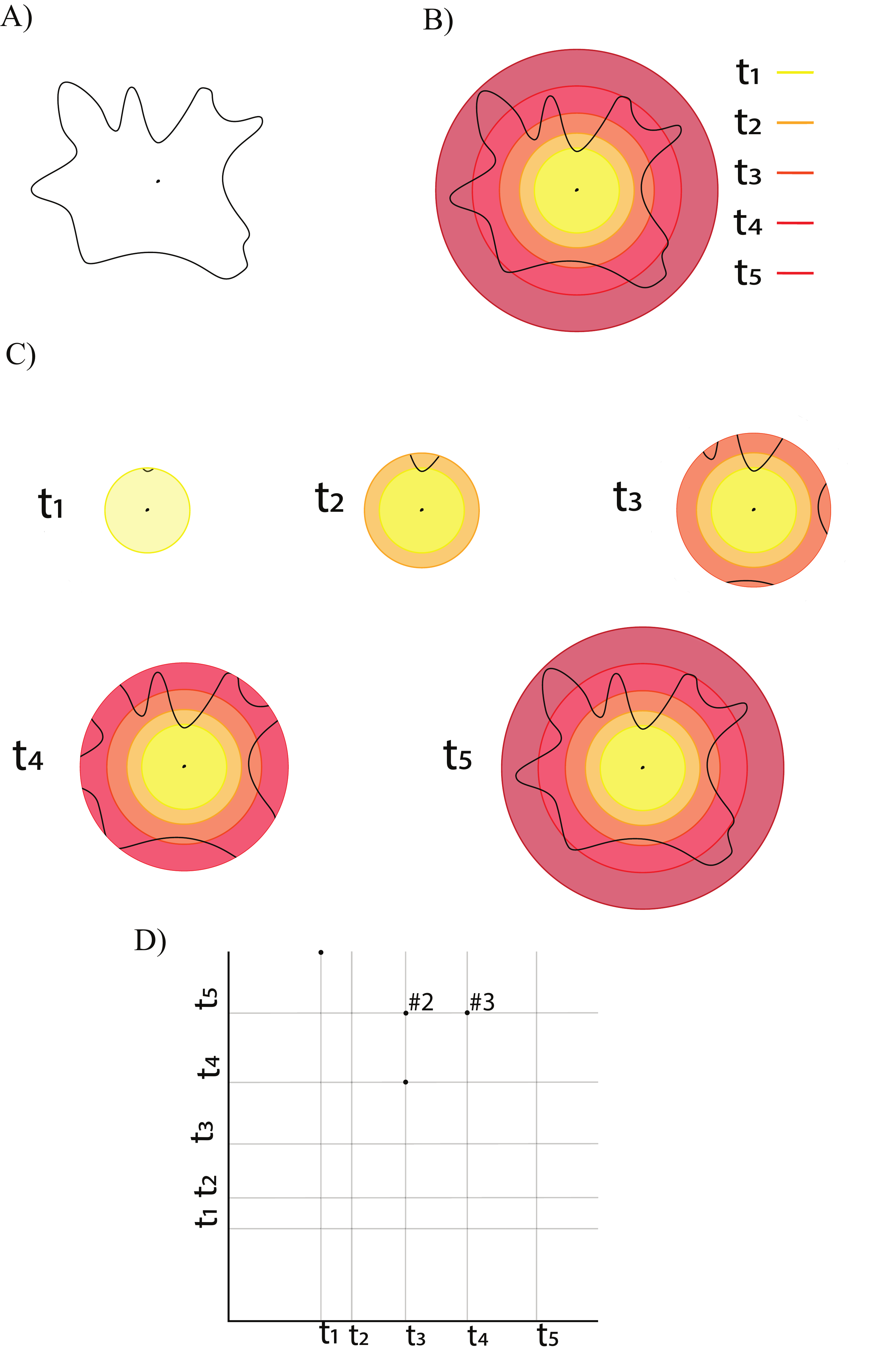}
            \caption{A) \textbf{\textit{The input data:}} a cell contour and the centre of its nucleus marked; the latter serves as the base point for the radial distance function. B) \textbf{\textit{The radial distance function:}} The complete cell contour forms a graph $G$. The edges of this graph are measured relative to the cell centre by computing the largest Euclidean distance between the centre and the endpoints of the edge:  the corresponding measure is the radial distance function with respect to the centre. Edges whose radial distance function is below a given cut-off value (or `time step'), illustrated as concentric circles around the centre, define a sub-graph of the whole contour. C) \textbf{\textit{Graph filtration:}} Examples of sub-graphs for five different time steps. The different graphs obtained at increasing values of time form a filtration of the graph $G$. D) \textbf{\textit{The persistence diagram}} captures the topological properties of the graph filtration. The points marked as `$\#2$' and `$\#3$' indicate that the corresponding points have multiplicity 2 and 3, respectively, in the persistence diagram.} \label{fig:example-radial}
        \end{center}
    \end{figure}

\subsection{Illustration on a simple cell contour}\label{ex:radial}
    
        The \textit{input contour} $C$ (see \Cref{fig:example-radial}A), with the centre of the nucleus marked, forms a graph. Using the centre as a reference point, we construct a \textit{radial distance function} to the graph as follows: for vertices, we use the standard Euclidean distance to the centre of the nucleus, and for edges, we take the maximum of the distances of their two endpoints. Vertices and edges whose radial distances are below a certain threshold (or `time step'), form a sub-graph of $C$ (\Cref{fig:example-radial}B). The \textit{persistence diagram }(\Cref{fig:example-radial}D), captures the changes in the connected components of the sequence or filtration of sub-graphs of $C$ obtained at increasing time values.

        \begin{remark}
            The definition of sub-level sets (\cref{def:sublevel}) is cell-wise constant, rather than piecewise-linear one. The distance of a point on an edge from the centre of the function is not the standard Euclidean distance in $\mathbb{R}^2$, but rather the maximum of the distances of the two vertices. This is not an issue, as the difference in these two values is bounded, and does not change the topological information we obtain.
        \end{remark}
        
        The relationship between the sequence of sub-graphs and the persistence diagram is as follows. At $t_1$, we see the birth of a single connected component, which has infinite life and corresponds to the point $(t_1, \infty)$ in the diagram (where $\infty$ is represented by being at the top of the diagram). At $t_2$, there are no changes (no birth or death events). At $t_3$, $3$ connected components are born. At $t_4$, a component born at $t_3$ merges with another component (and hence dies), which corresponds to the point $(t_3, t_4)$. We also see the birth of $3$ components. At $t_5$, we have a single connected component, formed by the remaining $2$ components born at $t_3$ merging with the component born at $t_1$, corresponding to the multiplicity $2$ point $(t_3, t_5)$, and all $3$ components born at $t_4$ merge with the original component as well, corresponding to the multiplicity $3$ point $(t_4, t_5)$.

        As a multi-set of points, the persistence diagram is
        \begin{align*}
            \mathcal{D} = \left\{ (t_1, \infty),  (t_3,t_4), (t_3,t_5), (t_3,t_5), (t_4,t_5), (t_4,t_5), (t_4,t_5)\right\},
        \end{align*}

        \noindent and, since we are only considering the connected components, we call this a \emph{dimension $0$} persistence diagram.

    As we are using graphs to represent each contour, we can also consider the information captured by the cycles in the sub-graph filtration. Each contour is a simple, closed curve in $\mathbb{R}^2$, and hence the corresponding graph $G$ contains a single cycle. Furthermore, this cycle appears only in the filtration when the \emph{last} vertex appears. While it is an important descriptor of the \emph{size} of the contour, it is inefficient to capture this information in a \emph{dimension $1$} persistence diagram. Hence, we modify our dimension $0$ diagram as follows, so that we capture this information: we pair the birth of the essential class in dimension $0$ with the birth of the essential class in dimension $1$. In this case, the set of points in the persistence diagram becomes 
    
    \begin{align*}
        \mathcal{D} = \left\{ (t_1, t_5),  (t_3,t_4), (t_3,t_5), (t_3,t_5), (t_4,t_5), (t_4,t_5), (t_4,t_5)\right\}.
    \end{align*}
    
    \begin{remark}
        Readers familiar with persistent homology and persistence diagrams will notice that this is a non-standard modification. Due to the nature of the contours, performing this \emph{essential pairing} allows us to more efficiently represent and compare the topological descriptors. Further, readers familiar with extended persistent homology will notice that this pairing is \emph{baby extended persistence} on the boundary curve.
    \end{remark}
    
    \FloatBarrier
 
\subsection{Comparing two persistence diagrams using the Wasserstein distance}\label{sec:wd}
    A persistence diagram provides a summary of the changes in the connected components as we progress along the sequence of graphs. Consider two sequences of graphs 
        
    \begin{equation*}
            \mathfrak{G}^1 = G_0^1 \rightarrow G_1^1 \rightarrow \ldots G_{\alpha_1}^1
    \end{equation*}
        
    and 
         
    \begin{equation*}
        \mathfrak{G}^2 = G_0^2 \rightarrow G_1^2 \rightarrow \ldots G_{\alpha_2}^2,
    \end{equation*}
        
    \noindent corresponding to two cell contours, with their associated persistence diagrams $D_1 = \mathfrak{D}(\mathfrak{G}^1)$, $D_2 = \mathfrak{D}(\mathfrak{G}^2)$. 
    We define the distance between the cell contours as the distance between $D_1$ and $D_2$, where the distance is the \emph{Wasserstein distance}, defined below.

    Imagine that there are $N$ farms that serve $N$ markets, and assume balance, that is, that each farm produces enough fruits and vegetables as needed by one market.
    A company in charge of the distribution of the produce from the farms to the market will take into account the individual cost of transport from any farm to any market to find an ``optimal transportation plan'', namely an assignment of farms to markets that leads to a minimal total cost for the transport.
    The seemingly simple problem can be traced back to the work of Monge in the 1780s \cite{Monge:1781}.
    What makes it so interesting is that its solution includes two essential components.
    First, it defines the assignment between farms and markets, enabling the registration between those two sets. 
    Second, and more relevant to us, it defines a distance between the set of farms and the set of markets, with such a distance being referred to as the Monge distance, the Wasserstein distance, or the earth mover's distance, depending on the field of applications.
    Formally, if $F$ is the set of farms and $M$ the set of markets, and if we define $C(i,j)$ the cost of transport between farm $i$ and market $j$, the assignment problem refers to finding a bijection $f$ between $F$ and $M$ that minimises
    \begin{equation}
    U = \sum_{i \in F} C(i,f(i)).
    \label{eqn:U1}
    \end{equation}
    Note that $f$ can be seen as a permutation of $\{1,\ldots,N\}$. As mentioned above, the optimal $U_{min}$ is a distance between $F$ and $M$.
    This is the distance that we use to compare two cell contours on the basis of their persistence diagram.

    As described above, a persistence diagram is defined by a set of points. Let $S_1$ (resp. $S_2$) be the set of points associated with $D_1$ (resp. $D_2$):
        \begin{eqnarray*}
        S_1 = \{X_1, \ldots, X_N\} \\
        S_2 = \{Y_1, \ldots, Y_N\}    
        \end{eqnarray*}
        
    We define the cost matrix $C$ as a power of the Euclidean distance, that is,
    
    \begin{eqnarray*}
     C(X_i, Y_j) = || X_i - Y_j ||^p
    \end{eqnarray*}
    The $p$-Wasserstein distance between $S_1$ and $S_2$ is then:
    \begin{eqnarray*}
        W_p(S_1,S_2) = \left ( \min_{f} \sum_{x_i \in S_1} || X_i - f(X_i) ||^p\right)^{1/p}
    \end{eqnarray*}

    The formalism defined above assumes that the two sets of points $S_1$ and $S_2$ considered have the same size, that is, there are as many points in $D_1$ as there are points in $D_2$. There is no reason why this is the case. In the more general case, $S_1$ contains $N_1$ points and $S_2$ contains $N_2$, with $N_1 > N_2$, without loss of generality. 
    This problem, however, can easily be reduced to the balanced case presented above by adding $N_1-N_2$ pseudo, or ``ghost'' points in $S_2$ that the two corresponding sets have the same cardinality. The distance between a point is $S_1$ and one of these pseudo-points can be chosen arbitrarily. One option is to position the ``ghost'' points on the diagonal of $D_2$.
    
    In the following, we will use the 2-Wasserstein distance to compare two cell contours via their persistence diagrams.

\section{Materials and Methods}\label{sec:materials-and-methods}

    \subsection{Human Mesenchymal Stem Cells}\label{subsec:hmsc}
    
    Adult human mesenchymal stem cells (hMSCs) were purchased from Lonza ($\#PT-2501$ Basel, Switzerland) and cultured in low glucose DMEM (Gibco, $\#1885-023$) supplemented with 10\% FBS (Sigma-Aldrich, Ref. $F7524$), and 1\% penicillin/streptomycin (Gibco, \#15140122) in regular tissue culture treated flasks (greiner Bio-One, $75 cm^2$, $\#658175)$ at $37^\circ$ C and $5.0\%$ CO$_2$. Cells were kept subconfluent at low density all the time and passaged and split every two or three days using pre-warmed trypsin / EDTA ($0.25\%$) (Sigma-Aldrich T4049) for an incubation time of $3$ min after a washing step with buffered phosphate saline (PBS) (Gibco, $\#14190144$). Cells were seeded on ibidi µ-Dishes ($35$~mm, high, ibiTreat, Cat.No: $\#81156$) at a density of $500$ cells cm$^{-1}$ to maintain a sufficient number of isolated cells for observation and grown for $24$ hours under identical culture conditions. The cells were then washed once with PBS and chemically fixed for $5$min in a $10\%$ solution of formaldehyde (Sigma-Aldrich, 252549) in PBS. The cells were then permeabilized with TritonX (Sigma-Aldrich, T 9284) and extensively washed with PBS.
    %
    %
    
    \subsection{HeLa cells}\label{subsec:heLa}
    
    HeLa cells (Leibniz Institute, DSMZ, ACC 57, RRID:CVCL\_0030) were cultured in Dulbecco’s Modified Eagle’s Medium (DMEM ,Sigma-Aldrich D5671) supplemented with $10\%$ fetal bovine serum (Sigma-Aldrich F7524) , $5\%$ sodium pyruvate solution (Sigma-Aldrich S8636) and $5\%$ penicillin/streptomycin (Sigma-Aldrich P4333). Cells were grown in T-25 flasks (BioLite No. 130189 Thermo Scientific, Germany) at $5\% CO_2$ at $37^\circ$C. Cells were split at $80\%$ confluence every two to three days using pre-warmed trypsin / EDTA ($0.25\%$) (Sigma-Aldrich T4049) incubation of $3$ minutes for detachment up to a maximum of $20$ passages per cell batch.
    
    Briefly,  $25$~mm round coverslips were treated with $0.03$~N NaOH solution at $90^\circ$C for 30 minutes. The treated coverslips were coated with 3-aminopropyltriethoxysilane (APTES, Sigma-Aldrich,440140) for 5 minutes followed by washing steps with PBS. The coverslips were then incubated with the solution of $0.5\%$ glutaraldehyde (Sigma-Aldrich, G7651) for 30 minutes and extensively washed with PBS.
    The coverslips were then treated with $0.4$~mM  sulfo-SANPAH (Thermo Scientific 22589-50MG) solubilised in $50$~mM HEPES buffer at a pH of 8.0. Subsequently, the treated coverslips were placed under a UV lamp at a wavelength of 365 nm for 15 minutes for cross-linker activation, followed by washing with HEPES solution. After this, the coverslips were coated with $50$ µg/ml concentration fibronectin (Gibco, 33010018) overnight at room temperature. 
    
    HeLa cells were seeded at a density of 5000 cells per well in a 6 well plate. After 24 hours, the cells were fixed in paraformaldehyde (PFA) $4\%$ (m/V) after 24 hours. The cells were then permeabilized with $0.5\%$ (V/V) Triton for 15 minutes and blocked for 30 minutes with $1\%$ bovine serum albumin (BSA) in PBS.

    Actin was stained with fluorescently labelled phalloidin Atto 550 (Atto Tec AD 550-81) for 15 minutes, followed by washing with PBS. The nucleus was stained with Hoechst 33342 (invitrogen H3570).

    \subsection{Fluorescence Staining of Cells}\label{subsec:stain}
    Both cell samples were fluorescently stained to reliably quantify the contours of the cell and the nucleus after fixation. Filamentous actin was stained using fluorescent Phalloidin-Atto $550$ (ATTO-TEC GmbH, AD $550-81$) to quantify its contour and the nucleus was visualised using the DNA selective Hoechst dye (Invitrogen, Hoechst $\#33342$). 
    
    \subsection{Unbiased Microscopy}\label{subsec:unbiased-mircoscopy}
    
    The fixed hMSCs  were imaged on an inverted fluorescence microscope (Zeiss AxioObserver, Oberkochen, Germany) using a 20x objective (Zeiss, Plan-Neofluar, 440340-9904) and recorded by a sCMOS camera (Andor Zyla, $4.2$P USB$3.0$) using two filter sets (blue emission (Zeiss Filterset 49) and red emission (AHF, F46-008)) for the stained nucleus and actin, respectively. Since the cellular outline is the major morphological parameter we wanted to assess, we ensured unbiased data acquisition of the samples by inspecting using the nucleus fluorescence channel first and only selecting cells that were isolated (no other nucleus in the field of view) and had a healthy-looking non-deformed nucleus. Multiple nuclei, ill-shaped nuclei, as well as any oddly shaped nuclei were excluded to avoid recording cell outlines from abnormal cells. Subsequently, the actin channel of the cell was recorded to complete the data set for each cell.

    HeLa cells were imaged on the same inverted fluorescence microscope after fixation using a 40x objective (Zeiss, LD Achroplan, 440865) and an Andor Sona-4BV6X sCMOS camera.
      
    \subsection{Image Processing and Contour Generation}\label{subsec:image-processing}

    We used the \href{http://filament-sensor.de}{FilamentSensor2.0} tool \cite{FS2.0} to perform image processing and extract the contour of each cell from the actin image. Here, we used the feature ``Include Area-Outline'' to export the contour from the binarised image of the cells. The cell centre (i.e.e the focal point used as a center for the radial function) was defined as the centre of mass of the nucleus, obtained from the aligned microscopy image after segmenting the nucleus in Fiji [29] using Otsu’s thresholding method.

    \subsection{Converting Pixel Measurements to Physical Units}\label{subsec:conversion}
    In the analysis of microscopy images, measurements are first recorded in pixels and need to be translated into physical units, usually microns, to be interpreted accurately. This translation is dependent on the spatial calibration of the imaging system, which is indicated by the pixel size—the actual length each pixel represents. The conversion formula is d = n × s, where d is the size in microns, n denotes the pixel measurement and s is the scale factor in micron per pixel. The scale factor is determined by the total magnification of the objective lens, intermediate optics, and the physical size of the camera sensor pixels. The scale factors are $0.3155$ and $0.1639$ for the $X1$ (i.e., the hMSCs)  measurements, and for $Y1$ (i.e., the HeLa cells) measurements, respectively.

    \subsection{Contour Analysis: computing the distance between 2 cells}\label{subsec:contour-analysis}

    After extracting the contour from each image and identifying the centre of the nucleus, we convert it to the graph representation $G$. Recall that every vertex in $G$ is of degree $2$, and $G$ contains a single cycle. Let $V=\{v_i\}_{i=1}^n$ be the set of vertices of $G$, ordered clockwise around the contour. Then every edge $e$ of $G$ is of the form $(v_i, v_{i+1})$, where $v_{n+1} = v_1$. Before we obtain our sequence of graphs $\mathfrak{G}$, we \emph{clean} our graph representation $G$ of $C$ by replacing any set of consecutive edges $\{(v_i, v_{i+1}), \ldots, (v_{j-1}, v_j)\}$ which are co-linear (having the same $x$ or $y$ coordinate) with the edge $(v_i, v_j)$ and removing the vertices $v_k$ for $i < k < j$. 
    
    \begin{remark}
        Consider a contour $C$, and let $G$ be the original graph representation and $G'$ the graph after it has been cleaned. As the metrics on the edges of $G, G'$ are defined as the maximum of the values at the $2$ vertices, the sequences of graphs $\mathfrak{G}$ and $\mathfrak{G}'$ generated by these metrics on $G$ and $G'$, respectively, will have different topological features. In particular, connected components may be born \emph{later}, by the removal of vertices that are closer to the base point of the radial distance function. These changes in values are bounded and therefore, by the stability of persistence diagrams \cite{wasserstein-stability}, the distance between the respective persistence diagrams is also bounded. Although it is possible to generate contours where this cleaning process leads to large bounds on the distance between the persistence diagrams, the geometric features that lead to this are not of concern in our application. Hence, we prioritise computational efficiency and proceed with the cleaned graphs.
    \end{remark}

    Working with the cleaned graph $G_{X}$ for each cell $X$, we filter $G_{X}$ (see \Cref{def:sublevel} and \Cref{ex:radial}), and obtain a persistence diagram $D_X$ (\Cref{def:persistencediagram}). Then we construct a distance matrix $M$, using the 2-Wasserstein distance between the persistence diagrams $D_X, D_Y$ as the distance between two cells $X,Y$.

    \subsection{Clustering cells based on their contour}
    \label{subsec:clustering}
    
    Clustering is the task of regrouping cells such that those that belong to the same group, referred to as a cluster,
    are more similar to each other than to those in other clusters.
    We use the 2-Wasserstein distance between the corresponding persistence diagrams as the \emph{similarity} score between a pair of cells, see \Cref{sec:wd}.
    The clustering of the cells is then performed using the agglomerative hierarchical clustering analysis, or HCA.
    The is a bottom-up approach in which each cell starts in its own cluster, and pairs of clusters are merged iteratively until all cells belong to the same cluster. The whole procedure defines a clustering tree. Although the distance between two cells is clearly defined above, a key element is to define the distance between two clusters.
    When two clusters A and B are sets of elements,
    the distance between A and B is then defined as a function of the pairwise distances between their elements. Four common choices of linkage are:
    \begin{itemize}
        \item \textbf{\textit{Average linkage}}: the distance between two clusters is the arithmetic mean of all the distances between the objects of one and the objects of the other:
            \begin{eqnarray*}
                d(A,B) = \sum_{a \in A} \sum_{b \in B}\frac{d(a,b)}{|A| |B|}
            \end{eqnarray*}
            where $|\cdot|$ stands for cardinality. Average linkage, also called UPGMA, is the default linkage for most HCA implementations. 
        \item \textbf{\textit{Single linkage}}: the distance between two clusters is the smallest distance between the objects in one and the objects in the other.
            \begin{eqnarray*}
            d(A,B) = \min \{d(a,b), a\in A, b \in B\}
            \end{eqnarray*}
        \item \textbf{\textit{Complete linkage}}: the distance between two clusters is the largest distance between the objects in one and the objects in the other.
            \begin{eqnarray*}
            d(A,B) = \max \{d(a,b), a\in A, b \in B\}
            \end{eqnarray*}
        \item \textbf{\textit{Ward's linkage}} accounts for the variances of the clusters to be compared. For a cluster $A$, the variance $SSE(A)$ is defined as:
            \begin{eqnarray*}
            SSE(A) = \sum_{a\in A} d (a, m(A))^2
            \end{eqnarray*}
            where d is the underlying distance used to compare two objects and $m(A)$ is either the centroid (if it can be computed) or medioid of the cluster (the medioid is the point in $A$ that has the least total distance to the other points in $A$). The Ward distance between two clusters $A$ and $B$ is then:
            \begin{eqnarray*}
            d(A,B) = SSE ( A\bigcup B) - (SSE(A)+SSE(B))
            \end{eqnarray*}
    \end{itemize}
    
    The choice of the linkage can have a significant influence in the clustering found by HCA: for example, simple linkage only looks locally at cluster distance, and as such may lead to elongated clusters, while reversely complete linkage will have a tendency to generate more compact clusters. There is no consensus as to which linkage scheme to use for a specific data set; this is, in fact, an active area of research.
    
    To avoid possible biases associated with the choice of linkage, we will use all four options in our analyses, performing HCA with \cite{scikit-learn}. However, this requires a way to compare the results of one option with those of the others. We chose our own concept of purity to perform such a comparison, which is defined as follows. Let $C_1$ be one cluster identified with HCA with a linkage method $L_1$. It is possible that $C_1$ may not be identified as its own cluster within the tree $T_2$ generated with another linkage method $L_2$. To assess how well $T_2$ recognises $C_1$, we follow the following algorithm:
    \begin{itemize}
        \item [1)] First, we choose a seed, $S_1$, i.e. an object that belongs to $C_1$. We initialise a list of objects $O=\{S_1\}$.
        \item [2)] We identify the leaf of $T_2$ corresponding to $S_1$, and add to the list $O$ the object that has the same parent $P_1$ in $T_2$ as $S_1$.
        \item [3)] We find the parent $P_2$ of $P_1$ and add to $O$ all objects that are in the sub-tree of $T_2$ starting from $P_2$. We then set $P_1 \leftarrow P_2$.
        \item [4)] We repeat step 3 until $O$ contains all objects in $C_1$
    \end{itemize}
    If the results with the linkage $L_2$ match exactly the results with the linkage $L_1$, $O$ will be equal to $C_1$. However, in general, $O$ will be bigger because it will include objects that are found by $L_2$ to be similar to objects in $C_1$ that were not identified by $L_1$. The \emph{purity} $P(C_1 /L_2)$ of $C_1$ with respect to $L_2$ is then defined as:
    \begin{eqnarray}\label{eqn:purity}
        P(C_1 / L_2) = \frac{ N-|O|}{N-|C_1|}
    \end{eqnarray}
    where $|\cdot|$ stands for cardinality and $N$ is the total number of objects. Note that $P$ is between 0 and 1. The closer $P$ is to one, the more consistent the two linkage strategies $L_1$ and $L_2$ are with respect to $C_1$.

    
    \subsection{Other methods for cell contour comparisons}
    \label{subsec:compare}
 
 We compare our method for computing the distance between two cell contours (which we will refer to as PD, for persistence diagram), with traditional distances such as an aspect ratio distance and a distance based on Fourier descriptors, as well as a more recent method based on elastic shape. We briefly describe those three methods below.
 
\subsubsection{A rotation- and translation-invariant aspect ratio distance.} \label{subsubsec:aspect}
Given a planar contour $C=\{(x_i,y_i)\}_{i=1}^N$, we first remove
translation by centering the points,
  \begin{eqnarray}
\tilde{C} = C - \frac{1}{N}\sum_{i=1}^N (x_i,y_i).
  \end{eqnarray}
Rotation invariance is obtained by applying principal component analysis (PCA)
to the covariance matrix of $\tilde{C}$.  Let $\lambda_1 \ge \lambda_2$ be the
eigenvalues of this covariance matrix.  These eigenvalues represent the
variances of the contour along the major and minor principal axes,
respectively.  The (rotation- and translation-invariant) aspect ratio $A$ of
the contour is then defined as
  \begin{eqnarray}
A = \sqrt{\frac{\lambda_1}{\lambda_2}} \ge 1.
  \end{eqnarray}
An aspect ratio of $A=1$ corresponds to an isotropic shape, while larger
values indicate increasing elongation of the contour.

\paragraph{Aspect-ratio distance.}
To compare the elongation of two contours with aspect ratios $A_1$ and $A_2$,
we define a distance
  \begin{eqnarray}
d(A_1, A_2) = \left|\log A_1 - \log A_2\right|.
  \end{eqnarray}
The logarithmic transform is preferable to a linear difference
$|A_1 - A_2|$ because the aspect ratio is inherently multiplicative:
a change from $A=2$ to $A=4$ represents the same geometric deformation as a
change from $A=4$ to $A=8$.  Using the logarithm therefore produces a
scale-symmetric measure of elongation, in which multiplicative changes in
aspect ratio correspond to additive differences in the metric.  The resulting
distance $d(A_1,A_2)$ is non-negative, symmetric, and equals zero if and only
if the two contours have identical elongation.

\subsubsection{A rotation- and translation-invariant distance based on Fourier descriptors.} \label{subsubsec:fourier}

Fourier descriptors (FDs) provide a compact frequency-domain representation of 2D closed curves. By treating a planar curve as a complex-valued periodic function, FDs capture the shape's geometric properties in a form amenable to classification and comparison.

\paragraph{Complex Representation of Curves}

A closed curve in the plane can be represented parametrically as a complex-valued function:
\begin{equation}
z(t) = x(t) + iy(t), \quad t \in [0, 1]
\end{equation}
where $x(t)$ and $y(t)$ are the Cartesian coordinates along the curve, and the parameter $t$ traces the contour from start to finish with $z(0) = z(1)$ (closed curve condition).

For a discrete contour with $N$ sampled points, we have:
\begin{equation}
z_n = x_n + iy_n, \quad n = 0, 1, \ldots, N-1
\end{equation}

\paragraph{Fourier Decomposition}

The Fourier descriptors are obtained via the discrete Fourier transform (DFT):
\begin{equation}
F_k = \frac{1}{N} \sum_{n=0}^{N-1} z_n \, e^{-2\pi i k n / N}, \quad k = -M, \ldots, M
\end{equation}
where $k$ is the harmonic index and $M$ is the maximum harmonic considered (typically $M \ll N$).

Note that the original curve can be reconstructed using the inverse DFT:
\begin{equation}
z(t) = \sum_{k=-M}^{M} F_k \, e^{2\pi i k t}
\end{equation}
where $t \in [0,1]$ is the normalized curve parameter.

\paragraph{Invariance Properties}

To enable meaningful shape comparison, FDs must be invariant to geometric transformations that preserve shape identity:
\begin{itemize}
\item [a)] \textbf{\textit{Translation Invariance}}

Translation of the curve by a vector $(a, b)$ adds a constant to $z(t)$:
\begin{equation}
z'(t) = z(t) + (a + ib)
\end{equation}

This affects only the DC component $F_0$. Translation invariance is achieved by:
\begin{enumerate}
    \item Computing the centroid: $z_c =\displaystyle  \frac{1}{N}\sum_{n=0}^{N-1} z_n$
   \item Centering the curve: $z_n \leftarrow z_n - z_c$
\end{enumerate}
After centering, $F_0 = 0$, and all other coefficients $F_k$ ($k \neq 0$) are translation-invariant.

\item [b)] \textbf{\textit{Starting Point Invariance}}
The choice of starting point $t = 0$ on the curve is arbitrary. If the starting point shifts by $\tau$, the curve becomes:
\begin{equation}
z'(t) = z(t + \tau)
\end{equation}

In the frequency domain, this introduces a phase shift:
\begin{equation}
F'_k = F_k \, e^{2\pi i k \tau}
\end{equation}

Starting point invariance is obtained by normalizing the phase of all coefficients relative to $F_1$:
\begin{equation}
\tilde{F}_k = F_k \, e^{-ik\theta_1}, \quad \text{where } \theta_1 = \arg(F_1)
\end{equation}

This ensures that $\tilde{F}_1$ is real and positive, removing the dependence on the starting point.

\item [c)] \textbf{\textit{Rotation Invariance}}

Rotation of the curve by angle $\alpha$ multiplies the complex representation by $e^{i\alpha}$:
\begin{equation}
z'(t) = e^{i\alpha} z(t)
\end{equation}

This affects all Fourier coefficients:
\begin{equation}
F'_k = e^{i\alpha} F_k
\end{equation}

Since rotation only changes the phase, rotation invariance is achieved by using the magnitudes:
\begin{equation}
E_k = |F_k|, \quad k = 1, 2, \ldots, M
\end{equation}

\end{itemize}

The magnitude descriptors $\mathbf{E}=\{E_1, E_2, \ldots, E_M\}$ are invariant to rotation, translation, and starting point.

\paragraph{An FD distance}
To compare two contours $C_1$ and $C_2$ with magnitude Fourier descriptors $\mathbf{E}^1$ and $\mathbf{E}^2$,
we define a distance
 \begin{eqnarray}
d(C_1, C_2) = \sqrt{ \sum_{i=1}^M (E_i^1 - E_i^2)^2}.
 \end{eqnarray}
 The FD distance is parameterized by $M$, the number of harmonics considered.
 
\subsubsection{The elastic distance} \label{subsubsec:elastic}
 The elastic distance, based on the Square-Root Velocity (SRV) framework, provides a mathematically rigorous way to compare shapes while allowing for elastic deformations. Unlike rigid alignment methods, elastic distance accounts for both differences in shape and differences in parameterization, making it particularly suitable for comparing curves that may be non-rigidly deformed versions of each other.

\paragraph{Square-Root Velocity Representation}

A planar curve $\mathbf{c}(t) = (x(t), y(t))$, $t \in [0,1]$, is represented by its Square-Root Velocity (SRV) function:
\begin{equation}
\mathbf{q}(t) = \frac{\dot{\mathbf{c}}(t)}{\sqrt{\|\dot{\mathbf{c}}(t)\|}}
\end{equation}
where $\dot{\mathbf{c}}(t)$ denotes the tangent vector. The SRV representation has the key property that the $\mathbb{L}^2$ metric on $\mathbf{q}$ corresponds to the elastic metric on the curve $\mathbf{c}$.

The original curve can be recovered from its SRV via integration:
\begin{equation}
\mathbf{c}(t) = \mathbf{c}(0) + \int_0^t \mathbf{q}(s) \|\mathbf{q}(s)\| \, ds
\end{equation}

\paragraph{Shape Space and Reparameterization}

Two curves may represent the same shape but be parameterized differently. A reparameterization is a diffeomorphism $\gamma: [0,1] \to [0,1]$ (smooth, invertible, monotone increasing with $\gamma(0)=0, \gamma(1)=1$). Under reparameterization, the SRV transforms as:
\begin{equation}
(\mathbf{q} \circ \gamma)(t) = \mathbf{q}(\gamma(t)) \sqrt{\dot{\gamma}(t)}
\end{equation}

The space of shapes is defined as the quotient space of curves modulo reparameterizations and rigid motions (translation and rotation).

\paragraph{Elastic distance}

The elastic distance between two curves $\mathbf{c}_1$ and $\mathbf{c}_2$ is defined as:
\begin{equation}
d_{\text{elastic}}(\mathbf{c}_1, \mathbf{c}_2) = \min_{\gamma \in \Gamma, \mathbf{R} \in SO(2)} \|\mathbf{q}_1 - \mathbf{R}(\mathbf{q}_2 \circ \gamma) \sqrt{\dot{\gamma}}\|_{\mathbb{L}^2}
\end{equation}
where:
\begin{itemize}
    \item $\Gamma$ is the set of all reparameterizations (diffeomorphisms)
    \item $\mathbf{R} \in SO(2)$ is a rotation matrix
    \item $\|\cdot\|_{\mathbb{L}^2}$ is the $\mathbb{L}^2$ norm: $\|\mathbf{q}\|_{\mathbb{L}^2}^2 = \int_0^1 \|\mathbf{q}(t)\|^2 dt$
\end{itemize}

This minimization finds the optimal alignment (rotation) and reparameterization that brings $\mathbf{c}_2$ closest to $\mathbf{c}_1$.

\paragraph{Invariance Properties}

The elastic distance possesses the following invariances:

\begin{itemize}
    \item \textbf{Translation invariance}: Achieved by centering curves at their centroids before computing SRV
    \item \textbf{Rotation invariance}: Built into the optimization via $\min_{\mathbf{R} \in SO(2)}$
        \item \textbf{Reparameterization invariance}: Built into the optimization via $\min_{\gamma \in \Gamma}$, making it robust to different tracing speeds or starting points
\end{itemize}

Full descriptions on how to compute the elastic distance is available in Refs \cite{Joshi:2007, Mio:2007, Srivastava:2011, Dogan:2015}.

\subsubsection{Implementations}
We have implemented the computation of the aspect ratio distance and the Fourier descriptor distance as Matlab scripts. The implementation
of the elastic energy is a much larger project. Instead of implementing our own version, we relied on the Matlab package fdasrv\_MATLAB \cite{Tucker:2014}, available at \url{https://github.com/jdtuck/fdasrvf_MATLAB}.

\subsection{Classical Multidimensional Scaling}
\label{subsec:MDS}

Classical multidimensional scaling (cMDS) computes an embedding of a set of $n$ objects into a Euclidean
space of dimension $p$ from a matrix of pairwise dissimilarities.  Given a
symmetric matrix $D = (d_{ij})$ of distances, cMDS assumes that the distances are
Euclidean and seeks points $x_1,\dots,x_n \in \mathbb{R}^p$ whose pairwise
Euclidean distances approximate $d_{ij}$.

\paragraph{Computing the embedding}
Let $D$ be the matrix of squared distances, and let $J = I_n - \frac{1}{n}\mathbf{1}\mathbf{1}^\top$
be the centering matrix.
cMDS forms the \emph{Gram matrix}
\begin{equation}
B \;=\; -\tfrac{1}{2} \, J \, D \, J.
\end{equation}
If the dissimilarities arise from an exact Euclidean configuration, then
$B = XX^\top$, where $X$ is the matrix whose $i$th row is the coordinate vector
of object $i$ in the embedding space.

The matrix $B$ is symmetric and is diagonalized as
\begin{equation}
B = V \Lambda V^\top,
\end{equation}
where $\Lambda = \mathrm{diag}(\lambda_1,\dots,\lambda_n)$ contains the
eigenvalues in decreasing order,
$\lambda_1 \ge \lambda_2 \ge \cdots \ge \lambda_n$,
and $V$ contains the corresponding orthonormal eigenvectors.
Coordinates in $\mathbb{R}^p$ are obtained by retaining the largest $p$ positive
eigenvalues:
\begin{equation}
X_p \;=\; V_p \Lambda_p^{1/2},
\end{equation}
where $\Lambda_p$ is the diagonal matrix of the top $p$ positive eigenvalues and
$V_p$ the associated eigenvectors (also called principal components).

The contribution or explained variance $C_p$ of the principal component $V_p$ is defined as:
\begin{equation}
C_p = \frac{\Lambda_p}{\sum_{i=1}^n \Lambda_i}
\label{eqn:var}
\end{equation}

\section{Results and discussion}
\label{sec:results}

   The result section serves three purposes:
   \begin{itemize}
   \item[a)] \textbf{\textit{Identification of subpopulations in a set of cells}}. We report how our method PH for comparing the shapes of cells can be used to identify unusual cells (i.e., outliers) and subpopulations within a cell population.
   \item[b)] \textbf{\textit{Assessing the performance of PH for clustering cells}}. We check how PH can distinguish two different cell types when the true partitioning is known.
   \item[c)] \textbf{\textit{Comparing PH with existing methods for comparing cell contours}}. We compare PH with an aspect ratio distance, a Fourier descriptor distance, and a elastic distance.
   \end{itemize}
   
     For all three tasks, we consider both human mesenchymal stem cells (hMSCs) and HeLa cells. 
    HeLa cells are a long-established, immortalised cancer cell line. They are a commonly used cell line for culturing in \textit{in vitro} conditions, showing relatively consistent growth patterns and morphologies within a population, despite some heterogeneity.
    hMSCs are primary cells, on the other hand, derived from bone marrow directly from patients. They are not immortalised or further altered and exhibit an inherent natural heterogeneity, both in morphology and function, due to their preparation and selection method.
     
    We start with analyzing a population of hMSCs, $X1$ (\Cref{subsec:X1}) and a population of HeLa cells, $Y1$ (\Cref{subsec:Y1}), with the experimental setup and analysis pipeline described in \Cref{sec:materials-and-methods}.

\subsection{Finding subpopulations in collections of cells}

 \subsubsection[X1]{$X1$: sub-populations of hMSCs}
 \label{subsec:X1}

\begin{figure}[H]
        \centering
        \includegraphics[width=0.5\linewidth]{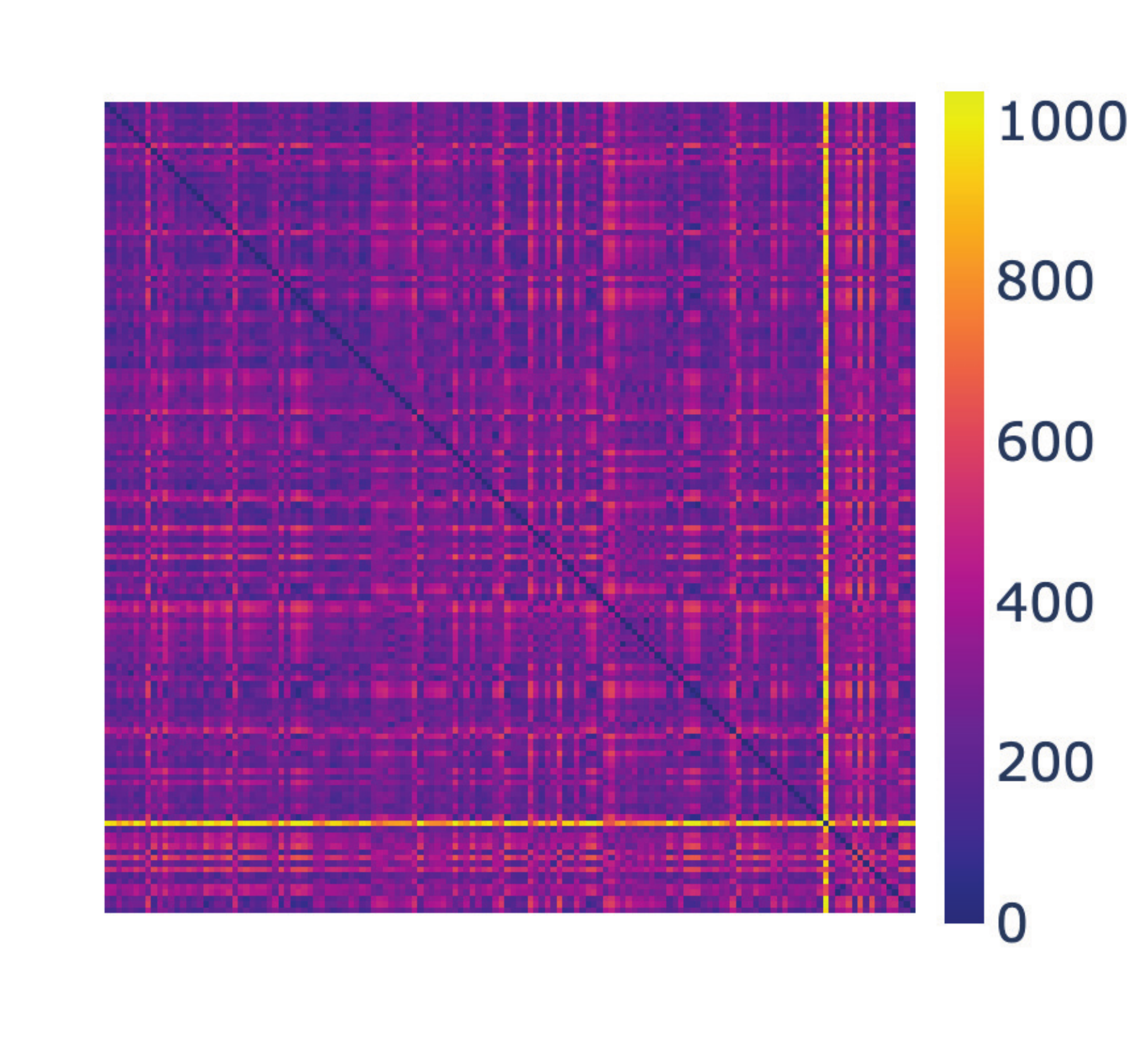}
        \caption{Heat map of the distance matrix for $X1$. There is a cell that has distinctly higher than average distances to the other cells, indicated by the row/column of mostly bright yellow. Generated with \cite{plotly}.}
        \label{fig:X1-heat-map}
    \end{figure}

\begin{figure}[H]
        \centering
        \includegraphics[height=3cm]{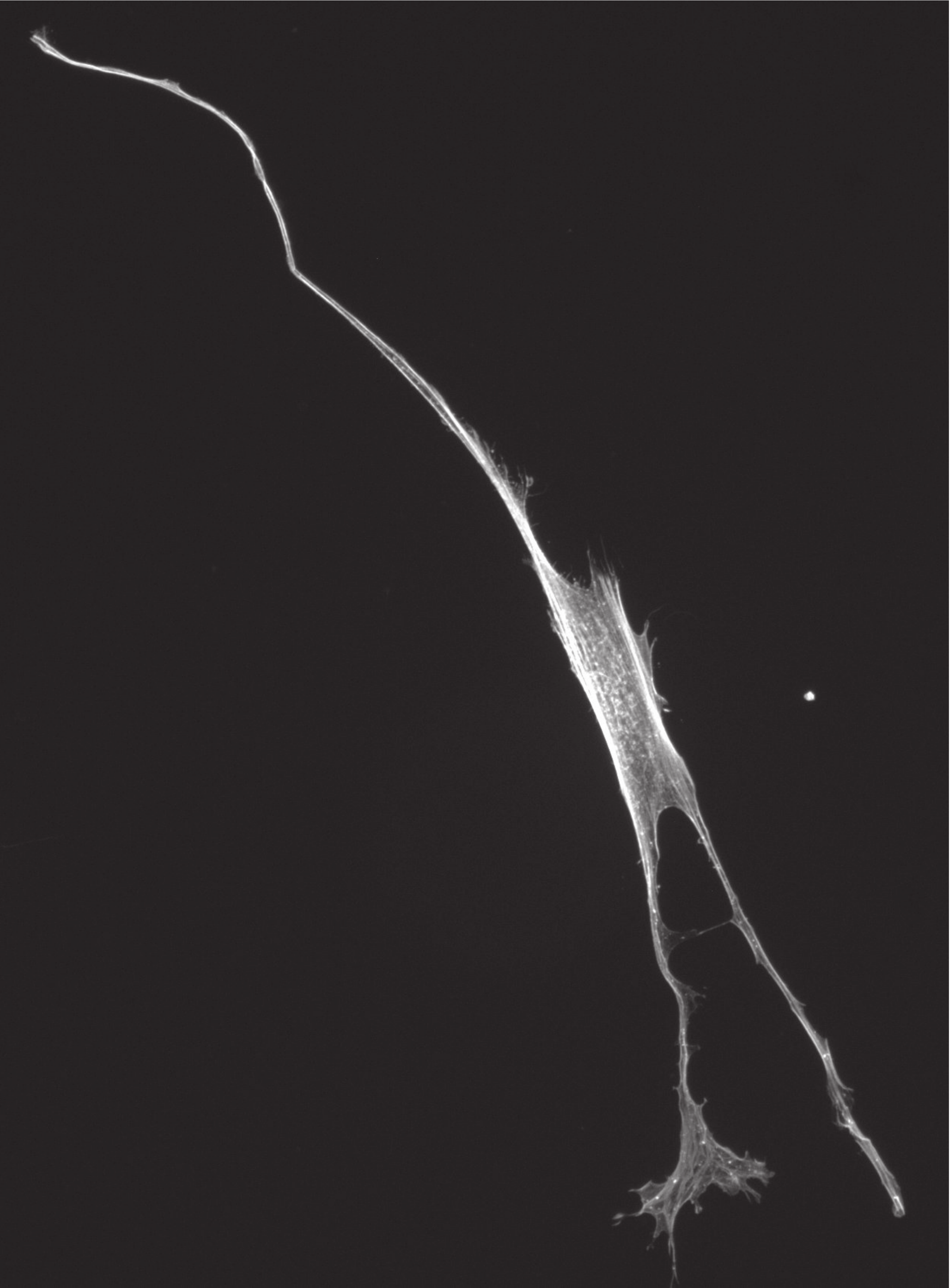}
        \caption{Image of the unusual cell shape (\texttt{X1-015}) identified in \Cref{fig:X1-heat-map} (image processed using \cite{ImageJ}.)}
        \label{fig:X1-015}
    \end{figure}

    The set $X1$ consists of $140$ cells. These cells have already been selected based on manual inspection, as described in \Cref{subsec:unbiased-mircoscopy}. To further analyse the homogeneity of this set of cells, we computed all pairwise distances between the cell contours using the persistence homology technique described above.  The corresponding distance matrix is visualised as a heat map in \Cref{fig:X1-heat-map}. The column/row of mostly bright yellow suggests that there is one cell that differs significantly from the others. This cell is shown in \Cref{fig:X1-015}.
    Clearly, this cell is oddly shaped: it is long and thin, with three long filipods, significantly different from the expected shape of an hMSC (see \Cref{fig:fig1} and \Cref{fig:X1-cluster-examples}).
    Such a shape is usually considered an outlier. 
    
\begin{figure}[h!]
       \centering
        \includegraphics[height=10cm]{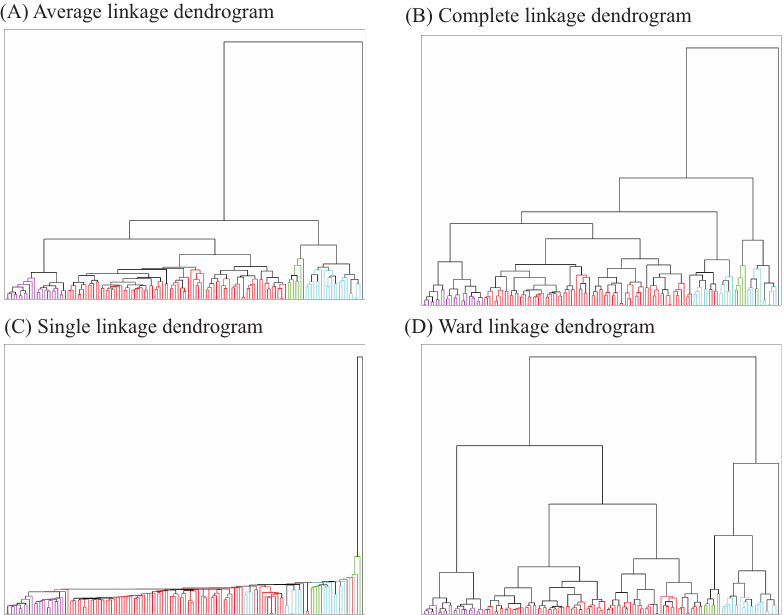}
        \caption{Dendrograms for $X1$, the colours correspond to 4 clusters obtained using average linkage. (A) Average linkage. (B) Complete linkage. (C) Single linkage. (D) Ward linkage.\\ In each of these, there is an outlier, with the corresponding leaf coloured purple. Generated with \cite{MATLAB}.}
       \label{fig:X1-dendrograms}
    \end{figure}

    We perform clustering of cell contours in $X1$ using the Wasserstein distance between their associated persistence diagram, in the presence (\Cref{fig:X1-dendrograms}), and the clustering in the absence of the `outlier' \texttt{X1-015} identified above is obtained by removing the single leaf branch on the right side of each dendrogram. We used HCA with four different linkages: average, complete, single, and Ward.

    As expected, cell \texttt{X1-015} is identified as its own cluster with all four linkages in \Cref{fig:X1-dendrograms} (it is the cell found at the extreme right of all four dendrograms). This cell has a unique shape that differentiates it from other hMSC cells. Although there are many possible reasons for this behaviour, \texttt{X1-015} is considered an outlier.     
    
    The clustering of the set $X1$ without the outlier \texttt{X1-015}, identifies subgroups. We obtain the dendrograms for $X1$ without \texttt{X1-015} by removing the singleton cluster in black which corresponds to \texttt{X1-015}. Those subgroups appear to differ under different choices of HCA linkage (\Cref{fig:X1-dendrograms}). This behaviour is not unexpected, as different linkage schemes capture different geometries for the cluster (see \Cref{subsec:clustering}). It is common to focus on only one linkage scheme, usually the average linkage, and ignore the others. Our approach is different. We use all four linkages and assess their consistency, as illustrated in \Cref{fig:X1-dendrograms}. We start with the average linkage scheme and cut the associated dendrogram to obtain four clusters. These four clusters are referenced as A (red, $86$ elements), B (blue, $7$ elements), C (green, $22$ elements), and D (purple, $24$ elements). We then consistently colour the dendrograms for all linkage schemes based on those clusters A, B, C, and D. As expected, there are differences. However, some consistencies are observed. For example, we note that cluster D (in purple) is grouped together across all $4$ linkage schemes. To confirm this visual consistency, we computed a purity score (see \Cref{eqn:purity}) of the clusters obtained with the average linkage in all four linkage schemes. The purity score quantifies how ``pure'' a group of objects is within a dendrogram. It is computed by first identifying the subtree within the dendrogram that contains all objects within that group. If this subtree only contains this group, it is deemed pure and the purity score is set to 1. If, instead, this subtree contains other objects, its purity is reduced. When the subtree is the whole tree, the purity score is reduced to 0. The purity scores of clusters A to D are reported in \Cref{tab:X1-main-purity}, while examples of cells for each clusters are shown in \Cref{fig:X1-cluster-examples}.

     \begin{table}[H]
        \centering
        \begin{tabular}{l|cccc}
                        & \multicolumn{4}{c@{}}{Cluster}\\ 
            Linkage     & A (red, $n=86$)   & B (blue, $n=22$) & C (green, $n=7$)& D (purple, $n=24$)\\ \hline
             average    & 1.000             & 1.000           & 1.000           & 1.000 \\
             complete   & 0.334             & 0.008           & 0.925           & 1.000 \\
             single     & 0.075             & 0.025           & 0.008           & 1.000 \\
             ward       & 1.000             & 1.000           & 1.00           & 1.000 \\
        \end{tabular}
        \caption{Purity score of the 4 clusters obtained with the average linkage for $X1$ without \texttt{X1-015}, see \Cref{fig:X1-dendrograms}. The colour and size of each cluster is in parentheses.}
        \label{tab:X1-main-purity}
    \end{table}

  \begin{figure}[H]
        \centering
        \includegraphics[width=0.8\textwidth]{fig6}
        \caption{Example cells from each cluster of the set $X1$ without \texttt{X1-015}. Those clusters are identified with HCA and average linkage scheme (see \Cref{fig:X1-dendrograms}). All cell images are shown at the same magnification level.  Images were processed using \cite{ImageJ}.}
        \label{fig:X1-cluster-examples}
    \end{figure}

 As mentioned above, cluster D (purple) is visually homogeneous within all four linkage schemes: this is confirmed as its purity scores remain equal to 1. The cells in this cluster have compact shapes and a prominent nucleus, as expected from cells that have been plated on glass. The same types of cells were distinguished as a sub-population FC by Haasters \emph{et al.} \cite{Haasters:2009}. In contrast, cluster A (red) is much less consistent within the different linkage schemes, with purity scores close to 0 (with the obvious exception of the average linkage). Visually, cells belonging to cluster A are more heterogeneous, with a star-shaped or a triangular shape (first row of \Cref{fig:X1-cluster-examples}). This group of cells maps with the sub-population RS identified by Haasters \emph{et al.}. Cells belonging to cluster B are significantly more elongated. Their purity score is high with the exception of the single linkage scheme, but this could just be anecdotal as there are only 7 cells in this cluster. They may correspond to spindle-shaped elongated cells, fibroblastic in shape, identified as SS cells by Haasters \emph{et al.}. The cells in cluster C are mostly compact, similar to those in cluster D, but usually larger. The purity scores of cluster C are close to 1, indicating that they form a group with homogeneous shapes. They were probably identified as belonging to the FC subpopulation by Haasters \emph{et al.}.

    \FloatBarrier

\subsubsection[Y1]{$Y1$: heterogeneity of HeLa cell shapes}\label{subsec:Y1}

\begin{figure}[H]
        \centering
        \includegraphics[width=0.5\linewidth]{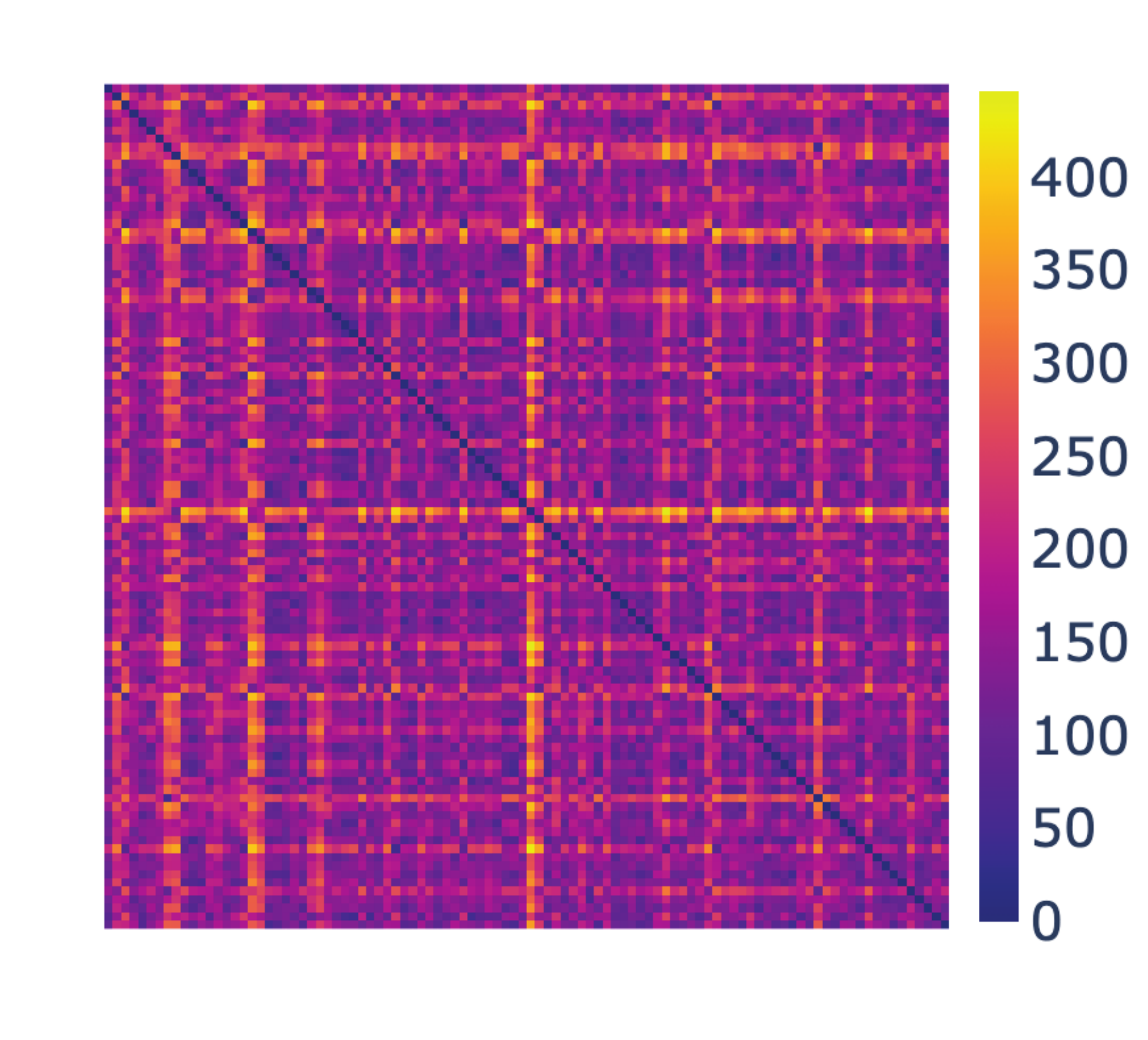}
        \caption{Heat map of the distance matrix for $Y1$. Generated with \cite{plotly}.}
        \label{fig:Y1-heat-map}
    \end{figure}

The set $Y1$ consists of $100$ cells. These cells have already been selected based on manual inspection; see \Cref{subsec:unbiased-mircoscopy}. We followed the procedure described in \Cref{subsec:X1} for the hMSCs to compute all pairwise distances between the cell contours using the methodology of \Cref{subsec:contour-analysis}. The corresponding distance matrix is visualised as a heat map in \Cref{fig:Y1-heat-map}, In contrast to the hMSCs for which we identified a clear outlier (see \Cref{fig:X1-heat-map}), we do not identify HeLa cells that we would unambiguously classify as outliers. 

\begin{figure}[H]
        \centering
        \includegraphics[width=\linewidth]{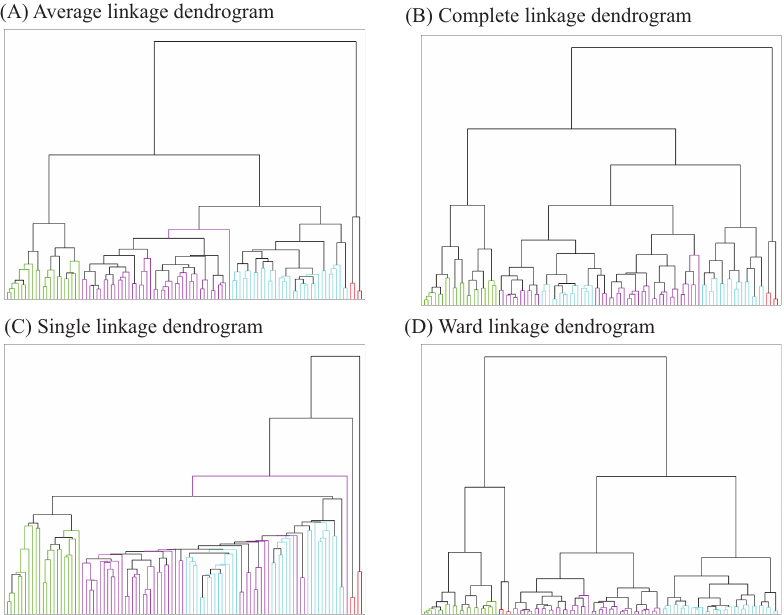}
        \caption{Dendrograms for $Y1$, the colours correspond to 4 clusters obtained using average linkage. (A) Average linkage. (B) Complete linkage. (C) Single linkage. (D) Ward linkage. In each of these, there is a consistent sub-population, coloured purple. Generated using \cite{MATLAB}.}
        \label{fig:Y1-dendrograms}
    \end{figure}

We clustered the cell contours in $Y1$ using the Wasserstein distance between their associated persistence diagrams (\Cref{fig:Y1-dendrograms}), using the four links from \Cref{subsec:clustering}. Similarly to the analysis of $X1$ above, we defined 4 clusters based on the average linkage scheme. These four clusters are referenced as A (red), with $7$ elements, B (green with $65$ elements, C (cyan, $15$ elements) and D (purple, $13$ elements). We then consistently colour the dendrograms for all linkage schemes based on those clusters A, B, C, and D. There are no consistencies in these 4 clusters obtained across the four linkages in \Cref{fig:Y1-dendrograms}. This is reflected in the purity scores that are not consistent (see \Cref{tab:Y1-purity}). This likely indicates that the HeLa do not separate into homogenous subpopulations. 

In \Cref{fig:Y1-cluster-examples}, we provide example cells for each of the four clusters obtained using the average linkage, cluster A (in red, $n = 4$ elements), cluster B (in green, $n = 33$ elements), C (in blue, $n = 21$ elements), and D (in purple, $n = 42$ elements). See \Cref{fig:Y1-cluster-examples} for plots of the contours with the centre mass of the nucleus marked. HeLa cells in all four clusters visually look homogeneous, with some shape heterogeneity;, i.e.with some cells having a more triangular structure, or some a pentagonal layout, and others are slightly more ellipsoidal. These three structures are present in each of the four clusters. Visually, the cells in cluster A are elongated, but there were only $4$ cells. In cluster C, the cells are more elongated and mostly have an ellipsoidal shape. Cluster D have a heterogeneous population of cells in terms of shapes, having a triangular structure, or some have a pentagonal layout.
In cluster B, the cells are comparatively compact (means have smaller values for cell area) with triangular or hexagonal shape. Overall, the HeLa cell population is homogeneous with variable shapes in each cluster, but no distinct identifiable sub-population is present. Taking into account the purity scores, we see that clusters A, B and D, have different purity scores across all types of linkage, suggesting the presence of heterogeneous shapes among these 3 clusters. HeLa cells are a well-established, robust cell line, and we imaged cells that all originated from the same cell culture flask at the same time point and identical experimental conditions. The presence of different clusters when analysing those cells, and the heterogeneity in those clusters revealed by differences in purity scores among different clustering techniques, all indicate that seemingly homogeneous cells that are expected to adopt similar conformations exhibit continuous variations in shapes.

     \begin{table}[H]
        \centering
        \begin{tabular}{l|cccc}
                        & \multicolumn{4}{c@{}}{Cluster}\\ 
            Linkage     & A (red, $n=4$)   & B  (blue, $n=33$)& C (green, $n=21$)& D (purple, $n=42$)\\ \hline
             average    & 1.000           & 1.000           & 1.000         & 1.000 \\
             complete   & 1.000           & 0.373           & 1.000         & 0.741 \\
             single     & 0.000           & 0.373           & 1.000         & 0.069 \\
             ward       & 1.000           & 1.000           & 1.000         & 1.000 \\
        \end{tabular}
        \caption{Purity score of the 4 clusters obtained with the average linkage for $Y1$, see \Cref{fig:Y1-dendrograms}. The colour and size of each cluster is in parentheses.}
        \label{tab:Y1-purity}
    \end{table}

  \begin{figure}[H]
        \centering
        \includegraphics[width=0.8\textwidth]{fig9}
        \caption{Example cells from each cluster of the set $Y1$. Those clusters are identified with HCA and average linkage scheme (see \Cref{fig:Y1-dendrograms}). All cell images are shown at the same magnification level.  Images were processed using \cite{ImageJ}.}
        \label{fig:Y1-cluster-examples}
    \end{figure}

\subsubsection[Z]{$X1$ vs $Y1$: two different types of variability}\label{subsec:Z}

The method we have developed follows a simple generic scheme when analysing a population: it computes distances between its members and uses those distances to define sub-groups within the whole population. Interestingly, the two examples described above, namely populations of hMSCs and of HeLa cells exhibit two different behaviours when investigated with our method, which illustrate intrinsic properties of those cell types.

Application of our method on hMSCs identified (at least) four different sub-populations in our sample. 
hMSCs are primary cells that are harvested from the bone marrow of patients and selected due to well-known surface markers. It is intrinsic to this procedure that hMSCs are a heterogeneous cell population, reflected in varying morphological and biological properties \cite{Settleman:2004, Delorme:2006, Li:2023}. It was established that an hMSC population can be divided into at least three subpopulations with intrinsic characteristics, small rapidly self-renewing cells, spindle-shaped cells, and large, flattened cells \cite{Smith:2004, Haasters:2009}. The consistent clustering results found with different HCA strategies is a good indicator of this discrete variability in the hMSC morphology.

In contrast, our clustering results indicate a more continuous variability for the morphology of the specific HeLa cell line. 
Those are immortalised cancer cells that are cultured for many passages \textit{in vitro} and have been reported to contain a very large number of genomic variants \cite{Frattini:2015}. A large study across different HeLa batches in many international laboratories identified substantial heterogeneity between HeLa variants \cite{Liu:2019}. However, for a specific type of this cell line, HeLa cells are expected to maintain a general epithelial-like polygonal morphology as a characteristic of the population (see \Cref{fig:Y1-cluster-examples}). Heterogeneity is then expected to reflect local environmental conditions, i.e. a continuous variability rather than a discrete set of sub-populations. This is well reflected in the absence of consensus in the different clustering strategies for the $Y1$ cell sample.

\subsection{Assessing the performance of PH: the dataset $X1Y1$}
\label{subsec:X1Y1}

The section above validated that the PH distance can be used to identify subpopulations in a set of single cell images. For the $X1$ dataset, we identified subgroups that agree with the previous study of Haasters \emph{et al.} \cite{Haasters:2009} while for the $Y1$ dataset, we found a more homogeneous set of cells, consistent with the expected behaviour of HeLa cells. None of these analyses, however, reflect the performance of PH as the true subpopulations of $X1$ or $Y1$ are not known.

\begin{figure}
        \centering
        \includegraphics[width=0.6 \linewidth]{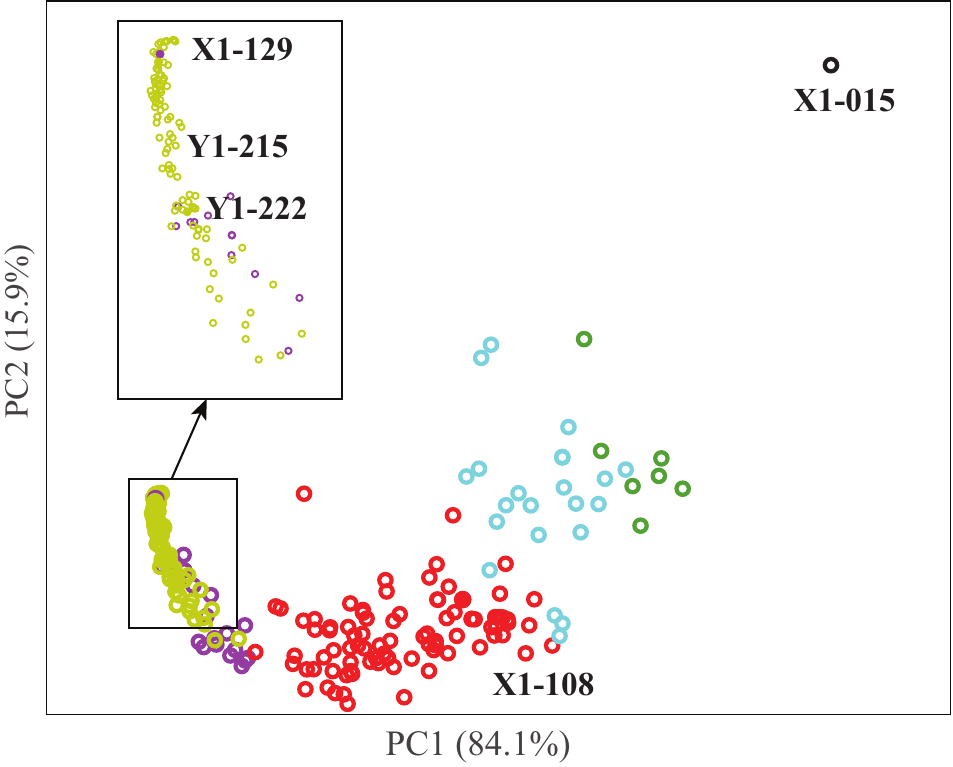}
        \caption{The MDS embedding of the combined dataset $X1Y1$. $Y1$ cells are represented in dark yellow; the region containing those cells is magnified in the insert. $X1$ cells are coloured according to the cluster they belong to, where the clusters have been defined in \Cref{subsec:X1}: cluster A in red, cluster B in blue, cluster C in green, and cluster D in purple. The outlier $X1-015$ (see \Cref{subsec:X1}) is shown in black.  The explained variance of each principal component is provided in parenthesis.}
        \label{fig:X1Y1-mds}
    \end{figure}

 It is possible to build a reference dataset by combining $X1$ and $Y1$ into $X1Y1$. We know that this set contains both hMS and HeLa cells, and we know for each cell which group it belongs to. Prior to analyzing this set, we made sure that all images of cells were at the same scale, using the translation from pixels to nm described in \Cref{subsec:conversion}.
  We followed the procedure described above to compute all pairwise distances between the cell contours of $X1Y1$ using the methodology of \Cref{subsec:contour-analysis}. The corresponding distance matrix was then subjected to multi dimensional scaling, as described in \Cref{subsec:MDS}.
  Results are presented in \Cref{fig:X1Y1-mds}. 
  
There are three important observations associated with  \Cref{fig:X1Y1-mds}:
\begin{itemize}
\item[a)] The $X1-015$ cell that was identified as an outlier when analysing $X1$ cells alone remains an outlier when analysing the combined dataset $X1Y1$.
\item[b)] HeLa cells and hMS cells are well separated, at least visually on the projected space. Note that there is one hMS cell that is found in the region of the HeLa cell, $X1-129$. This is discussed below.
\item[c)] The clustering of $X1$ cells  is mostly recovered when analysing the combined dataset $X1Y1$.
\end{itemize}
Those three observations hint at the robustness of our distance PH. These observations are qualitative, based on visual inspection of the MDS projected space. In the next section we will provide a more quantitative analysis.

  \begin{figure}[H]
        \centering
        \includegraphics[width=0.8\textwidth]{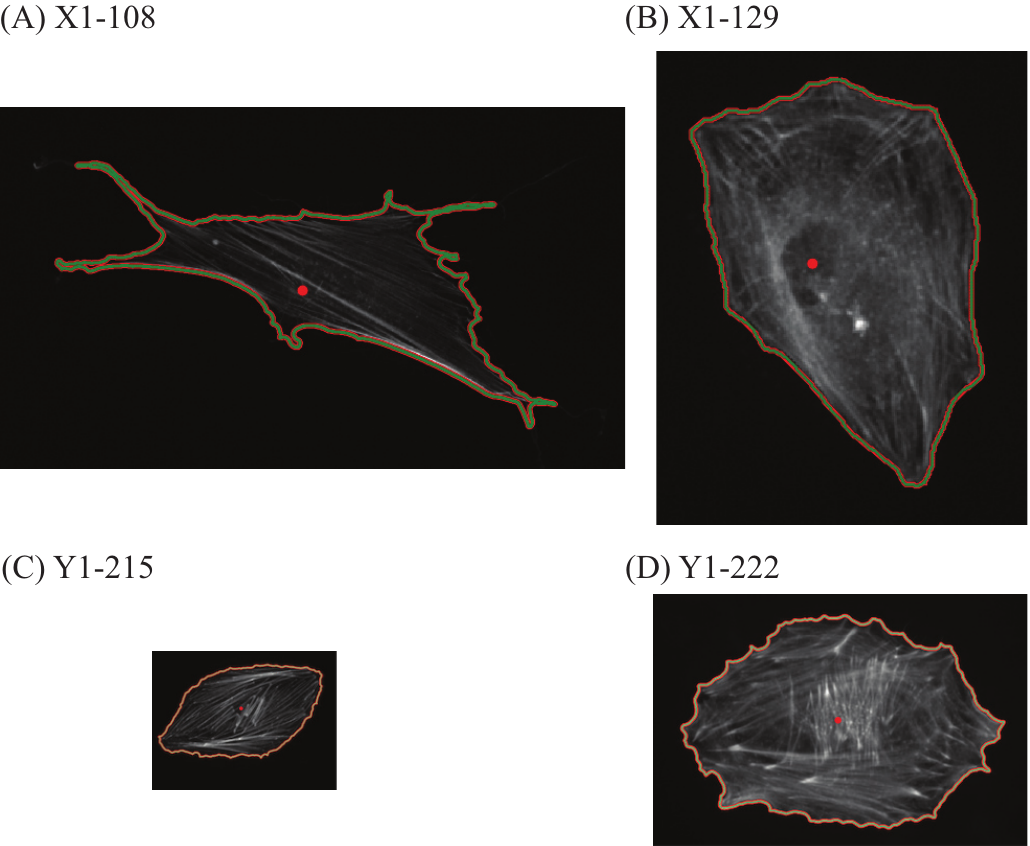}
        \caption{Example cells from  the set $X1Y1$. Those cells are identified in the MDS of $X1Y1$ in \Cref{fig:X1Y1-mds}.  All four cells are at the same scale. Images were processed using \cite{ImageJ}.}
        \label{fig:X1Y1-examples}
    \end{figure}

An interesting observation described above as the presence of one $X1$ cell in the space occupied by the $Y1$ cells. \Cref{fig:X1Y1-examples} compares this cell, $X1-129$, to another $X1$ cell, $X1-108$, and to two examples of $Y1$ cells, $Y1-215$ and $Y1-222$. We have observed earlier that $X1$ cells, namely hMS cells, are heterogeneous in shape when plating on glass. 
This is confirmed here when comparing the shape of $X1-108$, elongated with a few protrusions, with the shape of $X1-129$, rounder and smoother. $Y1$ cells, namely HeLa cells, have more homogeneous, round shapes, as illustrated with cells $Y1-215$ and $Y1-222$. It is therefore not unexpected that a round $X1$ cell is found similar to $Y1$ cells. Note that HeLa cells are smaller than hMS cells.

\subsection{Comparing the PH distance with other distances between cell contours}
\label{subsec:compdist}

    \begin{figure}
        \centering
        \includegraphics[width=0.65 \linewidth]{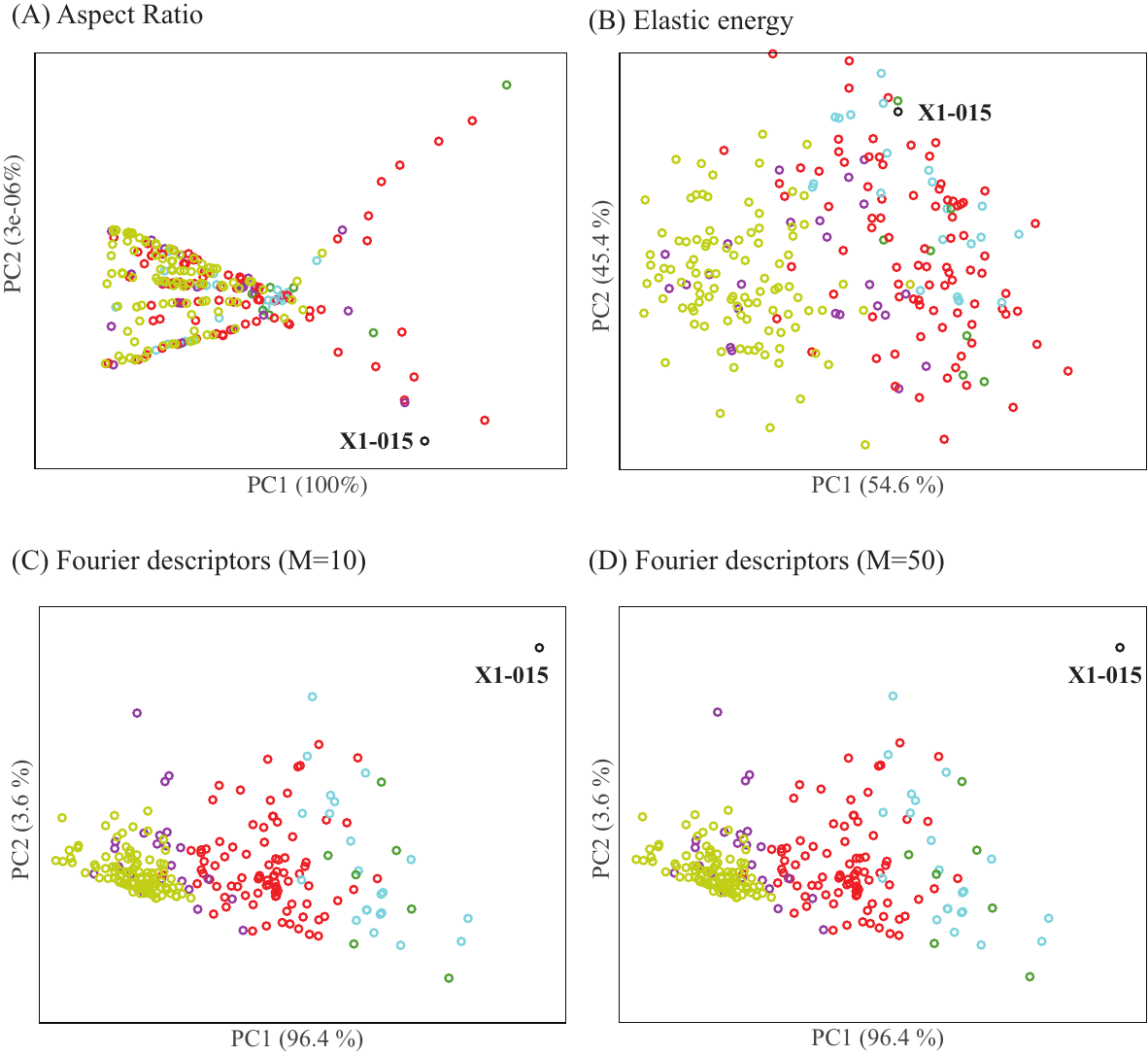}
        \caption{The MDS embeddings of the combined dataset $X1Y1$ for 4 different distances between cell contours, (A) aspect ratio distance, (B) elastic shape distance, (C) Fourier descriptors with $M=10$, and (D) Fourier descriptors with $M=50$ (see text for details on each of the distance). We label the position of $X1-015$, the known outlier among $X1$ cells, on all four panels. The explained variance of each principal component is provided in parenthesis. Colours are described in the caption of  \Cref{fig:X1Y1-mds}.}
        \label{fig:methods-mds}
    \end{figure}

In addition to PH, we tested 4 different approaches for computing the distance between two cells represented with their contours, namely closed 2D curves.
Each method is tested on their ability to distinguish HeLa cells ($Y1$) and hMS cells ($X1$) among the combined dataset $X1Y1$, as well as to recover the four clusters $X1_A$, $X1_B$, $X1_C$, and $X1_D$ among hMS cells.

The first approach compares cell shapes using a distance derived from their aspect ratios. Each contour is summarized by a rotation- and translation-invariant measure of elongation, capturing how stretched or isotropic the cell is. The distance between two cells is then defined by the absolute difference between their log-aspect-ratios, which reflects relative changes in elongation (see details in \Cref{subsubsec:aspect}. 

Our second family of methods characterizes each cell contour using Fourier descriptors, which capture shape information in the frequency domain. We test two resolutions: one using a small number of harmonics (M = 10) to capture coarse shape features, and another using a larger number (M = 50) to include finer geometric details. Distances between cells are computed as Euclidean distances between their descriptor vectors. 

The third approach uses an elastic shape metric that treats contours as parameterized curves and compares them through an alignment-based deformation framework. This method measures how much elastic bending and stretching is required to transform one contour into another, yielding a mathematically rigorous notion of shape similarity. Unlike the previous methods, which rely on global summaries or frequency coefficients, the elastic metric directly compares shapes in their continuous form and is sensitive to detailed geometric differences.
We computed all pairwise distances between the cell contours of $X1Y1$ using the four methods presented above. The corresponding distance matrices were then subjected to multi dimensional scaling. Results are presented in \Cref{fig:methods-mds}, and should be compared to the results in \Cref{fig:X1Y1-mds}. 

The aspect ratio distance is intentionally simple and aims to quantify broad geometric differences rather than fine structural variations. As such, it performs poorly on separating cells whose geometry are quite similar. This is  observed in \Cref{fig:methods-mds}A, with poor separation of HeLa cells and hMS cells. Even the outlier $X1-015$ is barely seen apart from the other cells.

The elastic shape distance performs better than the aspect ratio distance: HeLa cells separate from hMS cells, although there is still some non negligible overlap with $X1$ cells (see \Cref{fig:methods-mds}B). The four clusters of $X1$ cells, however, do not separate.

The Fourier-based distances perform quite well. The outlier $X1-015$ is well separated, so are the HeLa cells $Y1$ from the hMS cells $X1$.  Compared to PH (\Cref{fig:X1Y1-mds}), there is more overall between the clusters within $X1$. Interestingly, there are no big differences between the Fourier distance with 10 harmonics, \Cref{fig:methods-mds}C, and the Fourier distance with 50 harmonics, \Cref{fig:methods-mds}D. 

The discussion above is qualitative and based on visual inspection of the MDS projections of the cells. To provide a more quantitative assessment of the distances, we evaluate their performances using the receiver operating characteristic (ROC) analysis \cite{Fawcett:2006}. 
A pair of cells is defined as similar, or ``positive", if they belong to the same group, and ``negative" otherwise. 
Groups refer to two gold standards, level 1 and level 2. Level 1 refers to a gold standard for which cells either belong to $X1$ or to $Y1$. Level 2 refers to a finer model in which cells may belong to 6 groups, the four clusters of $X1$, A, B, C, or D, the outlier $X1-015$ as its own group, and $Y1$.
All cells in $X1Y1$ are then compared using one of the distances considered above. 
For varying thresholds of the distance, all pairs below the threshold are assumed positive, and all above it are negative. 
The pairs that agree with the gold standard (i.e. Level 1 or Level 2) are called true positives (TP), while those that do not are false positives (FP). ROC analysis compares the rate of TP as a function of the rate of FP; it is scored with the area below the corresponding curve, AUC. An AUC score of 1 indicates that all TP are detected first: this corresponds to the ideal measure. On the other hand, an AUC score of 0.5 corresponds to the first diagonal: TP and FP appear at the same rate, and the distance is not discriminative. In addition to the full area below the ROC curve, we consider a partial AUC score, pAUC10, computed up to a false positive rate of 0.1 (i.e., 10\%). This 

    \begin{figure}
        \centering
        \includegraphics[width=0.8 \linewidth]{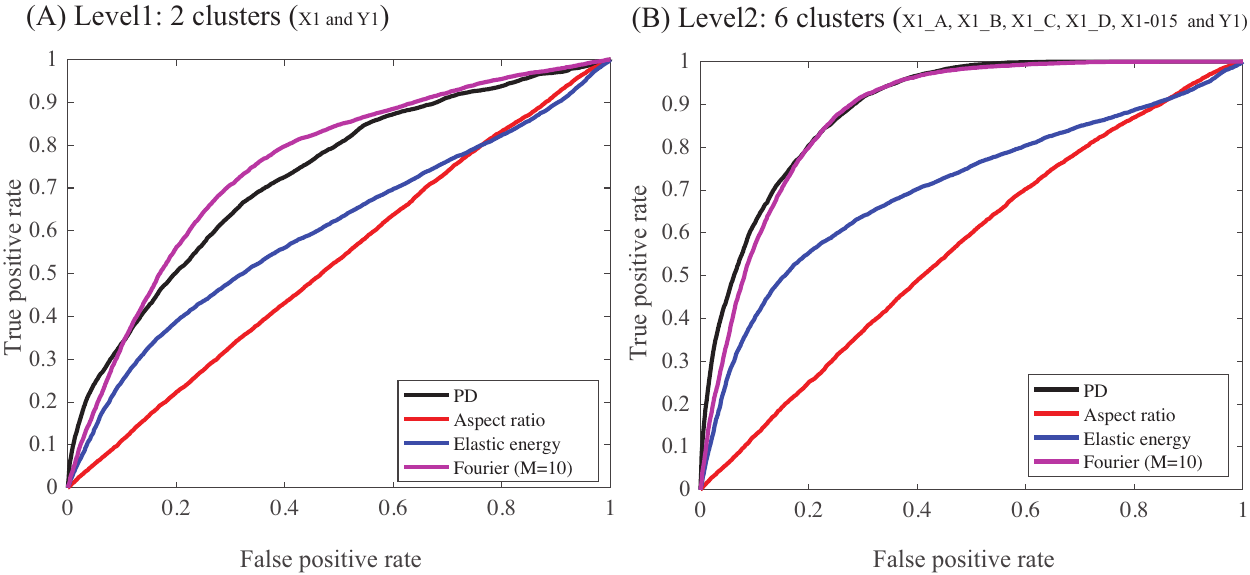}
        \caption{ROC analyses of four measures of cell geometry similarity. We compare the effectiveness of a the aspect ratio distance (in red), the elastic shape distance (in blue) the Fourier distance (with M=10) (in purple), and PH (in black) to measure cell contour similarities in the combined dataset $X1Y1$. “True” relationships are defined either with a binary separation of $X1$ and $Y1$ cells, Level 1 (panel A), or with a finer set of clusters involving the four clusters of $X1$, the outlier $X1-015$ as its own cluster, and $Y1$ (6 clusters, Level2, panel B). Curves close to the first diagonal (such as the ROC curve for the aspect ratio distance) indicate poor performance, while the upper most curve (such as the PH-based curve) indicates good performance.}
 \label{fig:roc}
    \end{figure}

 \begin{table*}[!hbt]
        
	\centering
        \begin{threeparttable}
        \caption{ROC analyses of the effectiveness of four measures of cell contour similarity in classifying cells in the combined $X1Y1$ dataset}
        \label{tab:ROC}
                \begin{tabular}{l c c c c c c}
                \cline{1-7} 
                & & \multicolumn{2}{c}{Level 1\tnote{a}} &  \multicolumn{2}{c}{Level 2 \tnote{a}} \\
                 \cline{3-4} \cline{6-7}
                Distanc measure & & AUC \tnote{b} & pAUC10 \tnote{c} & &  AUC \tnote{b} & pAUC10 \tnote{c} \\
                 \hline
                PH & &  0.73 & 0.03 && 0.89 & 0.05\\
                Aspect Ratio & & 0.53 & 0.006 && 0.57 & 0.006\\
                 Elastic distance & & 0.60 & 0.013 && 0.71 & 0.024\\
                 Fourier (M=10) & & 0.75 & 0.018 && 0.88 & 0.033 \\
                 Fourier (M=50) & & 0.75 & 0.018 && 0.88 & 0.033 \\
                                 \hline
                 \end{tabular}
                \begin{tablenotes}
                \item [a)] {\small Level 1 refers to a gold standard for which cells either belong to $X1$ or to $Y1$. Level 2 refers to a finer model in which cells may belong to 6 groups, the four clusters of $X1$, A, B, C, or D, the outlier $X1-015$ as its own group, and $Y1$. }
		\item [b)] {\small Area under the ROC curve; the higher the better.}
		\item [c)] {\small Area under the ROC curve integrated for false positive rate below 0.1. The higher the better.}
	         \end{tablenotes}
	         \end{threeparttable}
	         
\end{table*}

ROC analysis of the PH distance, Aspect Ratio distance, Elastic distance, and Fourier distance with 10 harmonics are shown in \Cref{fig:roc}A for level 1, and  \Cref{fig:roc}B for level 2. Note that we do not consider the Fourier distance with 50 harmonics as it was seen to lead to nearly identical results as the Fourier distance with 10 harmonics.
The corresponding AUC scores and partial AUC scores (pAUC10) are given in \Cref{tab:ROC}.

The ROC analyses confirmed the visual impression provided with the MDS projections of the cells, \Cref{fig:X1Y1-mds} and \Cref{fig:methods-mds}: aspect ratio distance only provides a base line, the elastic energy performs at an intermediate level between aspect ratio and PH or Fourier distance, while Fourier-based distances are close to the PH distance. Of interest, the ROC curves for the PH distance are very close to the y-axis (i.e. true positive axis) for up to 50\% true positive, much closer than the Fourier based distances. This is reflected in the pAUC10 values that are significantly better for PH than for the other distances.
This observation indicates that the PH distance is highly performant in the high sensitivity region, namely for short distances that are expected to detect similarity with high confidence. 

It does not escape us that the results on Level 2 are not void of biases. Indeed, the four clusters of $X1$ included in the gold standard have been defined with PH, i.e., with one of the methods that is tested. It remains that PH performs better than the other methods at Level 1; results at Level 2 should be considered as indicative of success.

\subsection{Parameter sensitivity: nucleus position}
\label{subsec:X1Y1}

All contour-comparison methods considered in the previous section share the same inputs, with the exception of PH, which additionally requires the location of the centre of mass of the nucleus. This point serves as the reference for the radial distance function underlying PH. Because the accuracy of this location may affect the resulting distances between contours, we assessed the sensitivity of PH to perturbations in the estimated centre of the nucleus. As described in the Methods section, the centre is obtained using Otsu thresholding applied to the segmented nucleus image.

To quantify this sensitivity, we conducted the following numerical experiments. For each cell in dataset $X1Y1$, the centre of the nucleus was independently perturbed by random shifts in both the $x$ and $y$ directions, each uniformly sampled from $[-s, s]$ pixels. If the perturbed centre falls outside the contour of the cell, the procedure is repeated. For each noise level $s$, all pairwise PH distances were recomputed and evaluated using the same ROC analysis as in the previous section. The relative root mean square (RMS) change in the distance matrix was defined as
\begin{eqnarray*}
\delta(s) = \frac{ | D(s) - D(0) |_F }{ | D(0) |_F },
\end{eqnarray*}
where $D(s)$ and $D(0)$ denote the distance matrices at noise levels $s$ and $0$, respectively, and $|\cdot|_F$ is the Frobenius norm.

\begin{figure}
        \centering
        \includegraphics[width=0.95 \linewidth]{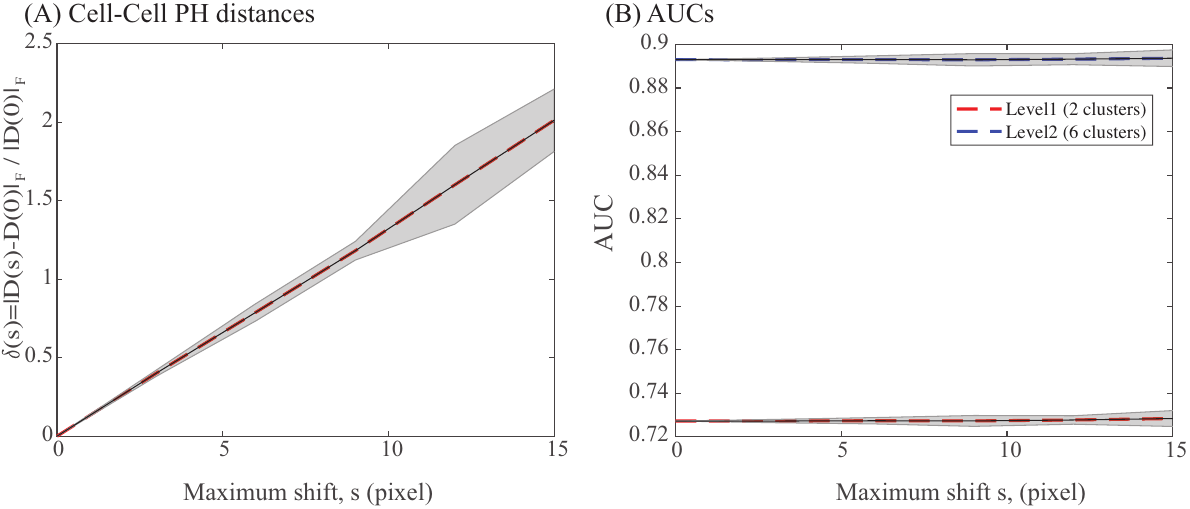}
        \caption{\textbf{Sensitivity of the PH distance to the position of the nucleus centre used as a focal point to the radial function}. Pairwise distances between cell contours of the $X1Y1$ dataset were computed after perturbing the position of the nucleus randomly between $[-s, s]$, where $s$ is a noise level in pixel. The distances are then assessed using a ROC analysis and measured with an AUC value (see text for details). Those experiments are repeated 20 times at each noise level. Panel (A) reports the mean (dashed line) and standard deviation (shaded area) of the relative RMS change of the individual distance as a function of $s$, while panel (B) shows the changes in $AUC$ values (both means, dashed lines, and standard deviations, shaded areas) at two level of granularity versus $s$ (see text for details).}
 \label{fig:sensitivity}
    \end{figure}

ROC performance of the distance matrix was quantified by the area under the curve, $AUC(s)$, at two levels of class granularity: (i) two clusters ($X1$ and $Y1$) and (ii) six clusters ($X1_A$, $X1_B$, $X1_C$, $X1_D$, $X1$–015, and $Y1$), see above for details. 
These experiments were repeated 20 times at each noise level.
Figures~\ref{fig:sensitivity}A and~\ref{fig:sensitivity}B show both the mean values and standard deviations over those 20 repeats for $\delta(s)$ and $AUC(s)$ as functions of the perturbation amplitude $s$, respectively.

As expected, the RMS deviation of the distance matrix increases monotonically with the noise level. In contrast, the classification performance remains largely unaffected, with $AUC$ values showing minimal degradation for perturbations of up to 15 pixels. This indicates that PH is robust to relevant inaccuracies in the estimated nucleus centre, as it can be usually determined with an uncertainty of maximal 2 px, well below 15 px..

 \section{Conclusion}\label{sec:discussion}

 Cell biologists commonly study in parallel the morphology of cells with the regulation mechanisms that affect this morphology. In the case of stem cells, for example, the shapes they assumed when plated on a substrate with different rigidities are expected to define the morphological descriptors of mechano-directed differentiation. However, the heterogeneous nature of the cell population is a major difficulty when studying the cell shape based on images from images from digital microscopes. It is common to manually assess first all the images associated with a population of cells under study in order to identify ``outliers", i.e. cells with unusual shapes that raise questions on their nature (i.e. these cells could be associated with contamination) or on the presence of experimental artefacts. The purpose of the present study was to propose an alternative automated method to help with this manual assessment. We have developed a new method for analysing cell shapes that is based on three elements:
 \begin{itemize}
    \item \textit{\textbf{A description of cell shapes using persistence homology}}. The shape of a cell is defined by its contour and the position of its nucleus. We compute a filtration of the edges that define the contour, using the radial distance to the nucleus as a filter. This filtration is used to define a persistence diagram that serves as a signature of the cell contour.
     \item \textit{\textbf{A distance between two cells}}. This distance is the Wasserstein distance between the persistence diagrams of their contours.
     \item \textit{\textbf{A measure of homogeneity of cell subgroups}}. We perform hierarchical clustering on cell shapes using the distance defined above, with four different linkage schemes. We define a purity score for subgroups of cells within the dendrograms associated with the clustering. This purity score reflects homogeneity.
 \end{itemize}
 
 We have tested our method on hMSCs, that are known to be heterogeneous with the presence of well-characterised sub-population, and HeLa cells, whose morphological differences are associated with continuous variability. We have shown that it automatically identifies unusual cells that can then be deemed outliers or not, as well as sub-populations that are consistent with previous analyses of sub-populations of hMSCs \cite{Haasters:2009}, and that, in the case of HeLa, confirms the expected homogeneity.
 
 We have compared the performance of our PH distance with three other types of cell contour distance measures, namely an aspect ratio distance, an elastic shape distance, and a Fourier descriptor based distance on a dataset that combines hMS cells and HeLa cells. The results are promising: the PH distance performs well, better than the three other methods, especially in the high sensitivity regime, i.e. for short distances that are expected to capture well similarity. Much remains to be done, however,
 before PH distance is established as a robust measure of similarity between closed 2D curves. 

 A distance based on aspect ratio alone does not perform well: this is by no means a surprise as such a distance only captures global geometry and not the fine details of a cell contour.
 It does not mean, however, that it does not contain meaningful information.
 In fact, there are many morphometric parameters that could complement our PH distance and provide a more robust measure of similarity of cell contours.
 Parameters such as basic size and shape measurements (area, perimeter, width, and length), derived measures such as compactness, elongation, and convexity, and measure for differential geometry such as curvatures are all capturing information about the geometry of a cell. We will explore how to include such information in further studies focusing on comparing cell shapes.

In this study, our primary objective was to introduce the proposed method and to demonstrate its performance relative to existing measures of cell-contour similarity, using a limited number of datasets of modest size. While this allowed us to establish proof of principle and comparative effectiveness, a comprehensive sensitivity analysis of all components of the method was beyond the scope of the present work.

We did, however, examine the sensitivity associated with estimating the position of the centre of mass of the nucleus, a key parameter that defines the origin of the radial function underlying the PH-based distance. This analysis showed that the method is robust to moderate perturbations of this reference point. Nevertheless, several additional sources of variability remain to be systematically investigated. These include the procedures used to extract cell contours from raw microscopy images, the preprocessing steps applied to those contours (such as smoothing or resampling), and the choice of radial or filtration functions used to construct the persistence diagrams. Each of these design choices may influence the resulting distances and their biological interpretability.

More broadly, the end-to-end pipeline—from raw image acquisition to contour extraction, persistence computation, and distance evaluation—has not yet been fully automated. For widespread adoption by cell biologists, this pipeline should be robust, reproducible, and seamlessly integrated into existing image-analysis frameworks. Addressing these methodological and practical aspects, through larger-scale studies and more extensive sensitivity analyses, constitutes an important direction for future work. Our longer-term goal is to develop a fully automated and user-friendly tool that incorporates the proposed approach and facilitates its application to diverse biological imaging datasets.

 Finally, we note that this whole study is based on images of 2D projections of cells. Cells are ultimately 3D objects and should be studied as such.
 Unfortunately, there is limited data available on 3D images of single cells. 
 There is room, however, for methodology development associated with comparing those 3D geometric objects, including the extension of the methods described here to 3D surfaces. The elastic shape energy has been described and implemented for 3D objects \cite{Kurtek:2010, Jermyn:2012}; similarly, 3D Fourier descriptors have been proposed, for example based on spherical harmonics \cite{Kazhdan:2003, Medyukhina:2020}. We plan to develop similar extensions for our PH distance.



\section*{Data Availability}

The code for PH is available on Github \url{https://github.com/yossibokorbleile/correa} as well as an archive on Zenodo (\url{https://doi.org/10.5281/zenodo.10021222}). 

All images used in this paper, as well as the code to analyse them, are available on GitHub \href{https://github.com/yossibokorbleile/PDaMSoC-data}{PDaMSoC-data}, which is also archived on Zenodo (\href{https://doi.org/10.5281/zenodo.16760103}{10.5281/zenodo.16760103}). In the repository, there are 4 directories (\texttt{X1}, \texttt{X2}, \texttt{X3}, \texttt{Y1}) containing data sets, with \texttt{X1}, \texttt{X2}, \texttt{X3} consisting of populations of hMSCs and \texttt{Y1} a population of HeLa cells. In this paper, we presented analysis of the datasets \texttt{X1} and \texttt{Y1}. We did analyse \texttt{X2} and \texttt{X3}, but did not obtain results that were distinctly different from \texttt{X1}.

\section*{Acknowledgments}
    We thank Stephan Huckemann, Katharine Turner, Benjamin Eltzner, Stephan Tillmann, Fariza Rashid, Vanessa Robins, and Lamiae Azizi for many useful discussions at various stages of this project.  FR and PY gratefully acknowledge Matthias Weiss (Experimental Physics I, University of Bayreuth, Germany) for granting access to cell culture and laboratories, as well as funding consumables and the fruitful discussion that contributed to this work. We thank the two reviewers for the thoroughness of their reviews and qualities of their comments. This research was funded in whole or in part by the Austrian Science Fund (FWF) ESP9584724. For open access purposes, the author has applied a CC BY public copyright license to any author-accepted manuscript version arising from this submission.

\end{document}